\documentclass[aps,prl,reprint,preprintnumbers,amsmath,amssymb,longbibliography]{revtex4-2}
\usepackage{graphicx,bm,booktabs,tikz}
\usepackage{mathtools, mathrsfs}
\usetikzlibrary{arrows.meta,calc}
\usepackage[colorlinks=true,citecolor=blue!55!black,linkcolor=blue!55!black,urlcolor=blue!55!black]{hyperref}

\usepackage{anyfontsize}
\usepackage{pgfplots}
\pgfplotsset{compat=1.18}
\usetikzlibrary{plotmarks,calc}

\allowdisplaybreaks[4]

\newcommand{\cP}{\mathcal P}
\newcommand{\cC}{\mathcal C}
\newcommand{\cF}{\mathcal F}
\newcommand{\cM}{\mathcal M}
\newcommand{\one}{\mathbf1}
\newcommand{\ii}{\mathrm i}
\newcommand{\Tr}{\operatorname{Tr}}
\newcommand{\Iback}{\check{\mathcal I}_{B\to A}}

\begin{document}
\preprint{USTC-ICTS/PCFT-26-73}
\title{When Conformal Interfaces Become Logarithmic}
\author{Jin Chen}
\affiliation{Department of Physics, Xiamen University, Xiamen 361005, China}
\affiliation{Peng Huanwu Center for Fundamental Theory, Hefei, Anhui 230026, China}
\email{zenofox@gmail.com}
\date{\today}
\begin{abstract}
We construct conformal interfaces having defect conformal manifolds between two-dimensional rational CFTs, at least one of which is non-unitary. In a continuous family, interfaces with real and complex spectra can be marginally deformed to each other via their defect moduli parameters. An exceptional point then arises along the trajectory in the conformal manifold, where two interface operators with complex conjugate weights coalesce into a logarithmic pair. We reconstruct the entire trajectory of this novel spectral collision, independently, from self-annulus partition functions and bulk two point correlators in the presence of the interfaces. 
\end{abstract}
\maketitle

\emph{Introduction.---}
In critical systems, power laws are ubiquitous, with exponents determined by the scaling dimensions of operators. However, many observables in percolation, polymers and disordered critical systems exhibit interesting logarithmic behavior~\cite{Cardy:2013rqg}. These logarithmic operators, together with their ordinary partners, can furnish a Jordan block basis of the Hamiltonian~\cite{Gurarie:1993xq}. An exceptional point provides a natural mechanism for the emergence of the logarithmic operators: By tuning couplings of a system, the spectrum becomes degenerate and the corresponding operators coalesce, so that the Hamiltonian is no longer diagonalizable. Exceptional points always require complexified couplings in Hermitian systems, whereas a non-Hermitian one can admit exceptional points at its real couplings~\cite{Heiss:2012dx}.

In this Letter, we uncover a novel mechanism for the emergence of exceptional points and the associated logarithmic operators on conformal interfaces between two-dimensional rational CFTs, at least one of which is non-unitary. We construct a continuous family of conformal interfaces $\mathcal I_\lambda$ between them. The bulk CFTs are non-trivial interacting Hermitian systems with real spectra, whereas the conformal interface in between is parameterized by real marginal couplings $\lambda\equiv\{\lambda_i\}$ which span a defect conformal manifold $\mathcal M_\lambda$. Within this family, by varying the moduli parameters $\lambda$, one can interpolate from a totally reflective interface to one with nonzero energy transmission. The former consists of tensor products of Cardy states in the two bulk theories, and thus has a real spectrum. In contrast, the latter is generically a non-factorized conformal interface with a complex spectrum. Both of the interfaces have diagonalizable boundary Hamiltonians until they are joined through an exceptional point at $\lambda=\lambda_{\rm EP}$ in $\mathcal M_\lambda$, where two operators of complex conjugate weights coalesce into a logarithmic pair at the collision, and then evolve into two with real weights afterwards. Throughout the whole process, the $2d/1d$ coupled system maintains its conformality as the interface flow is marginal. At the exceptional point, the boundary Hamiltonian exhibits a rank-two Jordan block pattern in the basis of the logarithmic pair operators. 

\emph{Conformal interfaces from phantom currents.---}
Now we explain how to construct a conformal interface, between two bulk rational CFTs $\mathcal T_L$ and $\mathcal T_R$, with a defect conformal manifold. In the construction, non-invertible symmetries, realized in terms of topological defect lines (TDLs), will play a fundamental role. These symmetries provide not only a way to introduce marginal boundary operators on the interface~\cite{Antinucci:2025uvj}, but also a non-perturbative algebraic description of it~\cite{ChenRGCompanion}. For these reasons, we always require the two bulk CFTs to share a common set of TDLs that can transmit through the conformal interface $\mathcal I$. Throughout the paper, the TDLs considered are Verlinde lines which, together with their fusion and junction data, will furnish fusion categories $\mathscr C_L$ and $\mathscr C_R$ in the bulk theories $\mathcal T_L$ and $\mathcal T_R$, respectively~\cite{Frohlich:2004ef,Frohlich:2006ch,Bhardwaj:2017xup,
Chang:2018iay,Thorngren:2021yso,Chang:2022hud}. 

\begin{figure}[tbp]
\centering
\begingroup
\definecolor{PCLeftFill}{RGB}{247,226,226}
\definecolor{PCRightFill}{RGB}{226,238,249}
\definecolor{PCFoldFill}{RGB}{235,229,244}
\definecolor{PCWall}{RGB}{181,43,114}
\resizebox{\columnwidth}{!}{%
\begin{tikzpicture}[
  x=1cm,y=1cm,>=Stealth,
  panel edge/.style={draw=black!55,line width=0.45pt},
  boundary/.style={PCWall,line width=1.8pt},
  tdl/.style={black,line width=1.15pt,dash pattern=on 3pt off 1.8pt},
  process arrow/.style={->,black!75,line width=0.75pt},
  step label/.style={font=\footnotesize,fill=white,inner sep=1pt},
  theory label/.style={font=\small},
  defect label/.style={font=\footnotesize},
  wall label/.style={font=\footnotesize,align=center}
]
  \path[panel edge,fill=PCLeftFill] (-1,1) rectangle (0,-1);
  \path[panel edge,fill=PCRightFill] (0,1) rectangle (1,-1);
  \draw[boundary] (0,1)--(0,-1);
  \node[theory label] at (-.50,.58) {$\mathcal T_L$};
  \node[theory label] at (.50,.58) {$\mathcal T_R$};
  \draw[tdl] (-.72,0)--(.72,0);
  \fill (-.72,0) circle[radius=1.6pt];
  \fill (.72,0) circle[radius=1.6pt];
  \node[defect label] at (-.50,-.38) {$\mathcal L_a^L$};
  \node[defect label] at (.50,-.38) {$\mathcal L_a^R$};
  \node[wall label,anchor=north] at (0,-1.10) {interface};
  \draw[process arrow] (1.13,0)--
    node[step label,above=2pt]{fold}
    node[step label,below=2pt]{$\mathcal T_R$} (2.17,0);
  \path[panel edge,fill=PCFoldFill] (2.30,1) rectangle (4.30,-1);
  \draw[boundary] (4.30,1)--(4.30,-1);
  \node[theory label] at (3.30,.58) {$\mathcal T_A$};
  \draw[tdl] (2.70,0)--(4.30,0);
  \fill (2.70,0) circle[radius=1.6pt];
  \node[defect label,anchor=north] at (2.70,-.08) {$X_a$};
  \node[defect label] at (3.65,-.38) {$\widehat{\mathcal L}_a$};
  \node[wall label,anchor=north] at (3.70,-1.10) {boundary};
  \draw[process arrow] (4.43,0)--
    node[step label,above=2pt]{shrink}
    node[step label,below=2pt]{TDL} (5.47,0);
  \path[panel edge,fill=PCFoldFill] (5.60,1) rectangle (7.60,-1);
  \draw[boundary] (7.60,1)--(7.60,-1);
  \node[theory label] at (6.60,.58) {$\mathcal T_A$};
  \fill (7.60,0) circle[radius=1.6pt];
  \node[defect label,anchor=east] at (7.47,0) {$X_a$};
  \node[wall label,anchor=north] at (7.00,-1.10) {boundary};
\end{tikzpicture}%
}
\endgroup
\caption{Folding and shrinking a transmitted TDL with chiral endpoint $X_a$}
\label{fig:folded-phantom-current}
\end{figure}
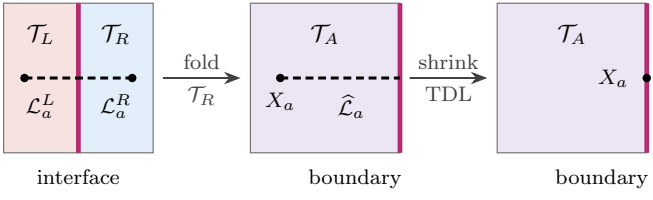

For a Verlinde line $\mathcal L_a^{L}\in \mathscr C_L$, and similarly for $\mathcal L_a^R\in\mathscr C_R$, one can always identify a twisted (anti-)chiral operator $\psi_a^L$ of conformal weight $(h_a^L,0)$ terminating on it. Since $\mathcal L_a^L$ is topological, one can shrink the line and attach the endpoint operator $\psi_a^L$ onto the interface~\cite{Gaiotto:2012np}. This observation is crucial to the existence of marginal boundary operators living on the wall $\mathcal I$. To see it, we employ the folding trick to consider the tensor product theory $\mathcal T_A=\mathcal T_L\otimes\overline{\mathcal T}_R$, where the overline denotes the orientation reversal. In this folded picture, the transmitted line then becomes a line ${\mathcal L}_a=\mathcal L_a^L\boxtimes\overline{\mathcal L_a^R}$ ending on $\mathcal I$ which now is a boundary state in $\mathcal T_A$. The chiral endpoint operator on $\mathcal L_a$ with chiral weight $h_a$ is given by the tensor product of a chiral $\psi_a^L$ and an anti-chiral $\overline{\psi_a^R}$, $X_a\equiv\psi_a^L\otimes\overline{\psi_a^R}$ with weight $h_a=h_a^L+h_a^R$. For integer $h_a$, the endpoints $X_a$ are known as ``phantom currents". The phrase ``phantom'' indicates that they are quasi-local operators attached to their defect lines $\mathcal L_a$. However, after shrinking $\mathcal L_a$ to the boundary $\mathcal I$, they become genuine local $1d$ operators therein. A spin-one phantom current $Q\equiv X_{a}$ of $h_a=1$ can then marginally deform the boundary state $\mathcal I$ to $\mathcal I_{\lambda}$~\cite{Antinucci:2025uvj,Furuta:2025ahl, Zhang:2026gqp}, with $\lambda$ the boundary coupling of $Q$, as shown in FIG.~\ref{fig:folded-phantom-current}.  

To utilize the spin-one phantom operator $Q$ to construct a continuous family of interfaces $\mathcal I_\lambda$, we will further introduce an auxiliary theory $\mathcal T_B$ by gauging the set of TDLs $\{\mathcal L_a\}$ with mutually local endpoints $X_a$. They therefore define a condensable Frobenius algebra $\mathcal L_\mathfrak a\equiv \oplus_a\mathcal L_a$ with $h_a\in\mathbb Z_{\geq 0}$, and give $\mathcal T_B=\mathcal T_A/\mathcal L_\mathfrak a$~\cite{Bhardwaj:2017xup,Fuchs:2002cm, Gaiotto:2020iye,Diatlyk:2023fwf}. This gauging is equivalent to condensing these phantom currents $X_a$ to the vacuum module, $\mathfrak A$, of $\mathcal T_A$ to form the vacuum module $\mathfrak B$ of $\mathcal T_B$, $\mathfrak B=\mathfrak A\oplus_a \mathcal X_a$, with $h_a\geq 1$, where $\mathcal X_a$ are the $\mathfrak A$-modules associated with $X_a$. Notice that the identity operator in $\mathcal T_A$ is the endpoint of the identity line, and thus is trivially a spin-zero phantom current. So $\mathcal T_B$ is a conformal extension from $\mathcal T_A$,
\begin{align}
    \mathcal T_A\xhookrightarrow{\ \mathfrak a\ } \mathcal T_B\,,\quad 
    \mathfrak B\big|_\mathfrak A=\mathfrak a\equiv \bigoplus_a \mathcal X_a\quad \text{for}\ \,h_a\in\mathbb Z_{\geq 0}\,.
\end{align}
In $\mathcal T_B$, the condensed spin-one phantom currents become local chiral currents $Q^i\in \mathfrak B$, with $i=1,\dots,d$ labeling the independent operators. Their zero modes $Q_0^i$ generate inner automorphisms $\Omega_\lambda$~\cite{Recknagel:1998ih, ChenRGCompanion}, 
\begin{align}
 U_{\lambda}=\exp\!\left(\ii\sum_{i=1}^d\lambda_i Q^i_0\right)\,,
 \qquad \Omega_\lambda=\operatorname{Ad}(U_\lambda)\,,
 \label{eq:general-inner}
\end{align}
where the $\lambda_i$ are real dimensionless parameters. Using $U_\lambda$, we can define a family of $\Omega_\lambda$-twisted Ishibashi states from an ordinary one $|\mu\rangle\!\rangle_{B}$ supported on a $\mathfrak B$-module $\mu$ in $\mathcal T_B$, $ \left|\mu;\lambda\right\rangle\!\rangle_{B}\equiv \left(U_\lambda\otimes\overline{\mathbf1}\right)\left|\mu\right\rangle\!\rangle_{B}$~\cite{Ishibashi:1988kg,Cardy:1989ir}. Obviously, they satisfy the $\Omega_\lambda$-gluing condition, $\left(\Omega_\lambda(W)_n-(-1)^h\overline W_{-n}\right)|\mu;\Omega\rangle\!\rangle=0$, where $W$ denotes a chiral current of weight $h$ in $\mathfrak B$. Since the inner automorphisms act within each module $\mu$ themselves, we can obtain a continuous family of Cardy states,
\begin{align}
    \left|\sigma;\lambda\right\rangle_{B}=\sum_\mu\frac{S^B_{\sigma\mu}}{\sqrt{S^B_{\mathbf v\mu}}}\,  \left|\mu;\lambda\right\rangle\!\rangle_{B}\,,
    \label{eq:Cardy_in_B}
\end{align}
where $S^B$ is the modular matrix of $\mathcal T_B$ and $\mathbf v$ denotes the vacuum label. For simplicity, we only consider the boundary state with brane label $\sigma=\mathbf v$, i.e. $|\mathbf v;\lambda\rangle_B$.

So far, $|\mathbf v;\lambda\rangle_B$ is a Cardy brane of the auxiliary theory $\mathcal T_B$. One has to further pull it back to a boundary state in $\mathcal T_A$ via a topological interface and its dual, defined by the Frobenius algebra $\mathcal L_\mathfrak a$,
\begin{align}
    &\mathcal I_{A\to B}:\ \mathcal T_A\ \longrightarrow\ \mathcal T_B\,,\qquad
    \check{\mathcal I}_{B\to A}:\ \mathcal T_B\ \longrightarrow\ \mathcal T_A\notag\\[3pt]
    {\rm with}\ \ &\mathcal I_{A\to B}\circ \check{\mathcal I}_{B\to A}=\mathcal L_\mathfrak a\,,\quad
    \check{\mathcal I}_{B\to A}\circ \mathcal I_{A\to B}=\mathcal L_\mathfrak b\,,
\end{align}
where $\mathcal L_\mathfrak b$ is the emergent quantum Frobenius algebra in $\mathcal T_B$, satisfying $\mathcal T_A=\mathcal T_B/\mathcal L_\mathfrak b$~\cite{Gaiotto:2020iye}. Fusing $\check{\mathcal I}_{B\to A}$ to the brane in~Eq.\eqref{eq:Cardy_in_B}, we can obtain a boundary state in $\mathcal T_A$ with a defect conformal manifold parameterized by $\lambda$,
\begin{align}
    \left|\mathcal I_\mathbf v;\lambda\right\rangle_A=\check{\mathcal I}_{B\to A}\,\left|\mathbf v;\lambda\right\rangle_{B}\,.
\end{align}
Unfolding $\mathcal T_A$, this boundary state $\left|\mathcal I_\mathbf v;\lambda\right\rangle$ corresponds to the proposed conformal interface $\mathcal I_\lambda$ between theories $\mathcal T_L$ and $\mathcal T_R$~\cite{Gaiotto:2012np,ChenRGCompanion}. 

The interface $\mathcal I_\lambda$ has the important feature that, for $\lambda=0$, it is simply the tensor product of Cardy states of $\mathcal T_L$ and $\mathcal T_R$, and thus totally reflective. On the other hand, when moving along the defect moduli to $\lambda\neq 0$, the stress tensors $T_L$ and $T_R$ will have non-trivial mixing with other spin-two phantom currents, leading to a non-zero energy transmission rate between $\mathcal T_L$ and $\mathcal T_R$. At a generic moduli loci, the interface $\mathcal I_\lambda$, or say $\left|\mathcal I_\mathbf v;\lambda\right\rangle$, is genuinely irrational with respect to the chiral algebra $\mathfrak A$. We refer the reader to~\cite{ChenRGCompanion} for full technical details. 

\emph{The model.---}
We begin with two simple non-unitary RCFTs where the spin-one phantom current is easy to construct. In the End Matter, a continuous family of conformal interfaces between the Ising CFT and the non-unitary minimal model $\mathcal M_{5,12}$ will be presented, where the models are better known but the interfaces are less obvious to construct.

Our ingredient is a class of simple affine superalgebras $\widehat{\mathfrak {osp}}(1|2)_k$ at level $k$, as a non-unitary generalization of $\widehat{\mathfrak{su}}(2)_k$ with two additional fermionic generators. The two bulk theories are given by cosets
\begin{align}
    \mathcal T_L=\frac{\widehat{\mathfrak {osp}}(1|2)_2}{\widehat{\mathfrak u}(1)_2}\,,\qquad
     \mathcal T_R=\frac{\widehat{\mathfrak {osp}}(1|2)_1\otimes\widehat{\mathfrak {osp}}(1|2)_1} {\widehat{\mathfrak {osp}}(1|2)_2}\,,
     \label{eq:osp}
\end{align}
with central charges $c_L=-\frac{3}{7}$ and $c_R=\frac{8}{35}$~\cite{Camino:1998pd}, as non-unitary parallels to parafermions and minimal models. In a more familiar language, $\mathcal T_L$ and $\mathcal T_R$ can be also given by the orbifolds of the following (non-)unitary minimal models,
\begin{align}
    \mathcal T_L=\frac{\text{Ising}\otimes\mathcal M_{4,7}}{\text{diag}\left(\mathbb Z_2^{\rm Ising}\times\mathbb Z_2^{\mathcal M_{4,7}}\right)}\,,\qquad
    \mathcal T_R=\mathcal M_{7,10}^{D\text{-type}}
    \label{eq:duality}
\end{align}
We summarize their more detailed modular data in the End Matter. The two bulks share a common fusion category of TDLs $\mathscr C=\{\one, x, y\}$ obeying the ${\rm SO}(3)_5$ fusion rules $ x^2=\one+x+y\,,\ xy=yx=x+y\,,\ y^2=\one+x$. Their endpoints $\{\psi^L_\one,\,\psi^L_x,\,\psi^L_y\}$ and $\{\psi^R_\one,\,\psi^R_x,\,\psi^R_y\}$ have (anti-)chiral conformal weights $\left(0,\,\frac{1}{7},\,\frac{3}{7}\right)$ in $\mathcal T_L$ and $\left(0,\,\frac{13}{7},\,\frac{4}{7}\right)$ in $\mathcal T_R$, respectively. So in the folded picture, we have three mutually local phantom sectors $\{X_1,\,X_x,\,X_y\}$ of spin-zero, two and one, respectively. The condensation leads to a conformal embedding,
\begin{align}
    &\mathcal T_A=\mathcal T_L\otimes \overline{\mathcal T}_R\xhookrightarrow{\,\mathfrak a\ } \mathcal T_B=\frac{\widehat{\mathfrak {osp}}(1|2)_1\otimes\widehat{\mathfrak {osp}}(1|2)_1} {\widehat{\mathfrak u}(1)_2}\,,\notag\\[2pt]    
{\rm with}\ \ &\mathfrak a\equiv \mathcal X_\one\oplus\mathcal X_x\oplus\mathcal X_y
\label{eq:phantom_sectors}
\end{align}
The spin-one phantom current $Q\equiv X_y$, after being condensed into the auxiliary theory $\mathcal T_B$, can be identified with the difference of the two Cartan currents in the numerator factors $\widehat{\mathfrak{osp}}(1|2)_1$, $J^z_1-J^z_2$, properly normalized as $Q(z)Q(0)\sim z^{-2}$. Using the zero mode of $Q$, $\Omega_\lambda=e^{i\lambda Q_0}$, and the ordinary Cardy state with identity label, $|b_\mathbf v\rangle_B$ in $\mathcal T_B$, we obtain an $\Omega_\lambda$-twisted Cardy brane,$|\mathbf v;\lambda\rangle_B=\sum_{\mu}\sqrt{S^B_{0\mu}}(U_{\mu,\lambda}\otimes\overline{\one})|\mu\rangle\!\rangle_B$, where $S^B$ is the modular matrix of $\mathcal T_B$, and $U_{\mu,\lambda}$ is the representation of $\Omega_\lambda$ on the ordinary Ishibashi state $|\mu\rangle\!\rangle_B$. We further apply the topological interface $\check{\mathcal I}_{B\to A}$ defined by the condensation algebra $\mathfrak a$, to transport $ |\mathbf v;\lambda\rangle_B$ to the physical product theory $\mathcal T_A$, $\left|\mathcal I_\mathbf v;\lambda\right\rangle_A=\check{\mathcal I}_{B\to A}\,\left|\mathbf v;\lambda\right\rangle_{B}$.

The brane $\left|\mathcal I_\mathbf v;\lambda\right\rangle_A$, after unfolding, is a genuine non-factorizable simple conformal interface $\mathcal I_\lambda$ parametrized by a real modulus $\lambda\sim\lambda+2\pi$. To see it, we compute the energy transmission coefficients, defined by the normalized one-point correlator of stress tensors $T_i\,\overline T_j$ in the presence of $\left|\mathcal I_\mathbf v;\lambda\right\rangle_A$ in $\mathcal T_A$~\cite{Quella:2006de},
\begin{align}
    K_{ij}(\lambda)=\frac{{}_A\langle T_i\,\overline{T}_j\big|\,\mathcal I_\mathbf v;\lambda\rangle_A}{{}_A\langle 0\big|\,\mathcal I_\mathbf v;\lambda\rangle_A}\,,\qquad 
    i,j=L,R\,.
\end{align}
The average transmission coefficient then reads as
\begin{align}
    \mathcal T(\lambda)=2\frac{K_{LR}+K_{RL}}{c_L+c_R}=\frac{160}{49}(\cos 2\lambda-\cos \lambda)\,.
\end{align}
Clearly, at $\lambda=0$ the boundary state is factorized as it is the pullback of the untwisted identity brane in $\mathcal T_B$, and thus completely reflective, i.e. $\mathcal T(0)=0$. On the other hand, for $\lambda=\pi$, $\mathcal T(\pi)=\frac{320}{49}>1$ is maximized, implying that the transmission coefficient in non-unitary CFTs would not be interpreted as a probability. However, a nonzero mixed stress tensor correlator already indicates that the interface allows energy transmission and is thus not factorized.

\emph{Spectral collisions on the defect conformal manifold.---}
Now we study the spectrum of the brane $\left|\mathcal I_\mathbf v;\lambda\right\rangle_A$. We will give numerical evidence that the brane is simple for $\lambda\neq 0\pmod {2\pi}$, whereas, at $\lambda=0$, it is consisted of 3 simple branes, each of which is a tensor product state of $\mathcal T_L$ and $\overline{\mathcal T}_R$. The spectrum can be extracted from the self-annulus partition function of $\left|\mathcal I_\mathbf v;\lambda\right\rangle_A$, whose closed and open channels are given by
\begin{align}
\label{eq:physical-annulus}
Z_\lambda(q)
 &=\left\langle \mathcal I_\mathbf v;\lambda\right|
 \widetilde q^{\,(L_0+\bar L_0-c_A/12)/2}\left|\mathcal I_\mathbf v;\lambda\right\rangle\\[4pt]
 &=\Tr_{\mathcal H_{\mathcal I_\lambda,\mathcal I_\lambda}}q^{L_0-\frac{c_A}{24}}=
 \chi_\mathbf v^B(q)+\sum_{\ell}n_\ell(\lambda)\chi^{\rm Vir}_{h_\ell(\lambda)}(q)
 \,,\notag
\end{align}
where $c_A=c_L+c_R=-\frac15$, $\widetilde q$ and $q$ are closed/open channel modular parameters, and $\chi_\mathbf v^B=\chi_\one^L\chi_\one^R+\chi_x^L\chi_x^R+\chi_y^L\chi_y^R$ is the vacuum character of $\mathcal T_B$ independent of $\lambda$. We compute the closed channel partition function from Eq.~\eqref{eq:physical-annulus}, and numerically extract the open channel spectrum of $\mathcal H_{\mathcal I_\lambda,\mathcal I_\lambda}$~\cite{ChenRGCompanion}. 

\begin{figure}[tbp]
\centering
\begingroup
\def\FigThreeCanvasWidth{3.375}
\def\FigThreeCanvasHeight{3.60}
\def\FigThreeAxesLeft{0.6075}
\def\FigThreeAxesWidth{2.66625}
\def\FigThreeRealBottom{1.26}
\def\FigThreeRealHeight{1.422}
\def\FigThreeImagBottom{0.504}
\def\FigThreeImagHeight{0.558}
\def\FigThreeLegendShiftX{0}
\def\FigThreeLegendShiftY{0}
\def\FigThreeColumnOne{0.963}
\def\FigThreeColumnTwo{1.687}
\def\FigThreeColumnThree{2.620}
\def\FigThreeLegendTop{3.348}
\def\FigThreeLegendRowGap{0.180}
\def\FigThreeLegendLineLength{0.124}
\def\FigThreeLegendTextOffset{0.159}
\def\FigThreeCurveWidth{1.2bp}
\def\FigThreeCircleRadius{1.5bp}
\def\FigThreeDiamondRadius{2.96985bp}
\def\FigThreeStarRadius{4.25bp}
\def\FigThreeLabelFont{\fontsize{9.5bp}{11.4bp}\selectfont}
\def\FigThreeTickFont{\fontsize{10bp}{12bp}\selectfont}
\def\FigThreeLegendFont{\fontsize{8.6bp}{10.3bp}\selectfont}
\def\FigThreeEPBendX{0.256}
\def\FigThreeEPLeaderEndX{0.365}
\def\FigThreeEPLabelX{0.380}
\def\FigThreeEPLabelY{0.030}
\def\FigThreeEP{0.23100173140221161}
\def\FigThreeEPH{0.084768331296448243}
\definecolor{FigThreeBlue}{HTML}{0072B2}
\definecolor{FigThreeOrange}{HTML}{D55E00}
\definecolor{FigThreeGreen}{HTML}{009E73}
\definecolor{FigThreeCyan}{HTML}{56B4C0}
\definecolor{FigThreePurple}{HTML}{A04478}
\definecolor{FigThreeGold}{HTML}{AE8219}
\definecolor{FigThreeIdentity}{HTML}{666666}
\definecolor{FigThreeEPColor}{HTML}{BF3159}
\definecolor{FigThreeGuide}{HTML}{A3A3A3}
\definecolor{FigThreeLeader}{HTML}{777777}
\definecolor{FigThreeEPText}{HTML}{555555}
\definecolor{FigThreeBand}{HTML}{F3F3F3}
\definecolor{FigThreeCutoff}{HTML}{BBBBBB}

\pgfdeclareplotmark{figthreediamond}{%
  \pgfpathmoveto{\pgfpointpolar{90}{\pgfplotmarksize}}%
  \pgfpathlineto{\pgfpointpolar{180}{\pgfplotmarksize}}%
  \pgfpathlineto{\pgfpointpolar{270}{\pgfplotmarksize}}%
  \pgfpathlineto{\pgfpointpolar{0}{\pgfplotmarksize}}%
  \pgfpathclose\pgfusepathqfillstroke
}
\pgfdeclareplotmark{figthreefilledstar}{%
  \pgfpathmoveto{\pgfpointpolar{90}{\pgfplotmarksize}}%
  \pgfpathlineto{\pgfpointpolar{126}{0.38196601125\pgfplotmarksize}}%
  \pgfpathlineto{\pgfpointpolar{162}{\pgfplotmarksize}}%
  \pgfpathlineto{\pgfpointpolar{198}{0.38196601125\pgfplotmarksize}}%
  \pgfpathlineto{\pgfpointpolar{234}{\pgfplotmarksize}}%
  \pgfpathlineto{\pgfpointpolar{270}{0.38196601125\pgfplotmarksize}}%
  \pgfpathlineto{\pgfpointpolar{306}{\pgfplotmarksize}}%
  \pgfpathlineto{\pgfpointpolar{342}{0.38196601125\pgfplotmarksize}}%
  \pgfpathlineto{\pgfpointpolar{18}{\pgfplotmarksize}}%
  \pgfpathlineto{\pgfpointpolar{54}{0.38196601125\pgfplotmarksize}}%
  \pgfpathclose\pgfusepathqfillstroke
}

\pgfplotstableread[row sep=\\]{
u re im mr mi er ei \\
0 0 0 0 0 nan nan \\
0.01 9.507728531350451e-05 -2.411998114743049e-11 nan nan nan nan \\
0.02 0.00038065220050841167 -5.6957295360548656e-10 nan nan nan nan \\
0.029999999999999999 0.00085777221608095567 -1.2078562086158735e-10 nan nan nan nan \\
0.040000000000000001 0.0015282112978745558 -8.2399201229389144e-11 nan nan nan nan \\
0.050000000000000003 0.0023945314402931857 -4.8819999036950551e-11 0.0023945314402931857 -4.8819999036950551e-11 nan nan \\
0.059999999999999998 0.0034601618968339944 -1.0536771614310287e-11 nan nan nan nan \\
0.070000000000000007 0.0047295148228822933 -3.0635184356481968e-12 nan nan nan nan \\
0.080000000000000002 0.0062081437411312789 1.9827965667002097e-13 nan nan nan nan \\
0.089999999999999997 0.0079029573430206349 -6.4584904920795424e-13 nan nan nan nan \\
0.10000000000000001 0.0098225116316045125 -3.4545046122540678e-12 0.0098225116316045125 -3.4545046122540678e-12 nan nan \\
0.11 0.011977409871199128 -2.4592767657690918e-10 nan nan nan nan \\
0.12 0.014380863952436897 -6.1866508688645078e-11 nan nan nan nan \\
0.13 0.017049488657394755 -3.486687586392133e-11 nan nan nan nan \\
0.14000000000000001 0.020004465510754427 -5.0160975881444799e-10 nan nan nan nan \\
0.14999999999999999 0.023273315265705603 -1.3115937488052255e-09 0.023273315265705603 -1.3115937488052255e-09 nan nan \\
0.16 0.026892659190303658 -2.3074928947190631e-10 nan nan nan nan \\
0.17000000000000001 0.030912862443730375 -1.6193942247821984e-10 nan nan nan nan \\
0.17999999999999999 0.035406234065950455 -6.0504796099640484e-11 nan nan nan nan \\
0.19 0.040482982953450306 -8.6515522454208508e-12 nan nan nan nan \\
0.20000000000000001 0.046326322249874825 -7.7722868766818105e-12 0.046326322249874825 -7.7722868766818105e-12 nan nan \\
0.20250000000000001 0.047941548724411968 -9.3943910394242205e-12 nan nan nan nan \\
0.20499999999999999 0.049632691035845632 -1.2104850047291011e-11 nan nan nan nan \\
0.20749999999999999 0.051409823531090845 -1.684578830922646e-11 nan nan nan nan \\
0.20999999999999999 0.053285769224182038 -2.5942054158847774e-11 nan nan nan nan \\
0.21249999999999999 0.055277284018429851 -4.5500964907119643e-11 nan nan nan nan \\
0.215 0.057406992453278266 -8.8813725098689355e-11 nan nan nan nan \\
0.2175 0.059706756921645357 -1.8456511031571757e-10 nan nan nan nan \\
0.22 0.06222403314407788 -2.0575804484465959e-10 nan nan nan nan \\
0.2225 0.065035339025302605 -8.8395887707725221e-11 nan nan nan nan \\
0.22500000000000001 0.068279728207177287 -1.586283841466115e-10 nan nan nan nan \\
0.22750000000000001 0.072268851623272498 -3.7359164488272134e-10 nan nan nan nan \\
0.23000000000000001 0.078168546166649311 -1.3842612638365981e-09 nan nan nan nan \\
0.23019999999999999 0.078873919888276953 -1.6557772740614447e-09 nan nan nan nan \\
0.23039999999999999 0.07967163990668813 -2.056973912703625e-09 nan nan nan nan \\
0.2306 0.080613601265482393 -2.7344336291086888e-09 nan nan nan nan \\
0.23080000000000001 0.081833274881953716 -4.2679577186328425e-09 nan nan nan nan \\
0.23089999999999999 0.082688682016745618 -6.4244562873609739e-09 nan nan nan nan \\
0.23094999999999999 0.08328768689887435 -9.4220275711958659e-09 nan nan nan nan \\
0.23097999999999999 0.083810014941093247 -1.5097647512866903e-08 nan nan nan nan \\
0.23099 0.084064707281583162 -2.0922472162679042e-08 nan nan nan nan \\
0.23100000000000001 0.084498337847740981 -5.6125855742652256e-08 nan nan nan nan \\
0.23100999999999999 0.084769264852532084 -0.00058958446076848401 nan nan nan nan \\
0.23102 0.084770434138327341 -0.00087636909585915644 nan nan nan nan \\
0.23105000000000001 0.084773923394555528 -0.001424552330768433 nan nan nan nan \\
0.2311 0.084779731800458155 -0.0020326954959987804 nan nan nan nan \\
0.23119999999999999 0.084791348231865196 -0.0028875495839684453 nan nan nan nan \\
0.23139999999999999 0.084814593583937767 -0.0040932231311150624 nan nan nan nan \\
0.2316 0.084837858604054628 -0.0050176492778206622 nan nan nan nan \\
0.23180000000000001 0.084861143892886917 -0.0057969754612020308 nan nan nan nan \\
0.23250000000000001 0.084942803542299702 -0.0079466280245598202 nan nan nan nan \\
0.23499999999999999 0.085236500415922165 -0.013009201142602103 nan nan nan nan \\
0.23749999999999999 0.085533417431293529 -0.016619995953833233 nan nan nan nan \\
0.23999999999999999 0.085833575642213569 -0.019598360862803411 nan nan nan nan \\
0.24249999999999999 0.086137098948599763 -0.022200013318108811 nan nan nan nan \\
0.245 0.086442973245553556 -0.024545196062451119 nan nan nan nan \\
0.2475 0.086752943672487579 -0.026701269366740277 nan nan nan nan \\
0.25 0.087065959329277048 -0.028710625210219744 nan nan nan nan \\
0.2525 0.087382012674327506 -0.0306020507454745 nan nan nan nan \\
0.255 0.087701895781099243 -0.03239637852681277 nan nan nan nan \\
0.25750000000000001 0.088024486862191928 -0.034108422583329717 nan nan nan nan \\
0.26000000000000001 0.088350294144393521 -0.035749923812057667 nan nan nan nan \\
0.27000000000000002 0.089686565959899822 -0.041771692710493975 0.089686565959899822 -0.041771692710493975 nan nan \\
0.28000000000000003 0.0910747650878099 -0.047161140484642088 nan nan nan nan \\
0.28999999999999998 0.092515235258504067 -0.052109521622635081 nan nan nan nan \\
0.29999999999999999 0.094008227263091035 -0.056727275133822253 nan nan nan nan \\
0.31 0.095553972402128234 -0.06108448611826247 0.095553972402128234 -0.06108448611826247 nan nan \\
0.32000000000000001 0.097152740080269084 -0.065228662013710775 nan nan nan nan \\
0.33000000000000002 0.098804810304519097 -0.069193502434775511 nan nan nan nan \\
0.34000000000000002 0.10051047946992199 -0.073003767620752474 nan nan nan nan \\
0.34999999999999998 0.10227006063451582 -0.076678148242347394 0.10227006063451582 -0.076678148242347394 nan nan \\
0.35999999999999999 0.10408388396132325 -0.08023105491048077 nan nan nan nan \\
0.37 0.10595229733830909 -0.083673786813739284 nan nan nan nan \\
0.38 0.10787566775062155 -0.087015323166173428 nan nan nan nan \\
0.39000000000000001 0.10985435923012803 -0.090262877233220587 nan nan nan nan \\
0.40000000000000002 0.11188882192772533 -0.093422285842796285 0.11188882192772533 -0.093422285842796285 nan nan \\
0.40999999999999998 0.11397944154737497 -0.096498312472297079 nan nan nan nan \\
0.41999999999999998 0.11612667020451216 -0.099494852540350537 nan nan nan nan \\
0.42999999999999999 0.11833097089393632 -0.10241510340074945 nan nan nan nan \\
0.44 0.12059281235262194 -0.10526166300494713 nan nan nan nan \\
0.45000000000000001 0.12291275744611278 -0.10803678275862733 0.12291275744611278 -0.10803678275862733 nan nan \\
0.46000000000000002 0.12529128024919406 -0.11074214229219767 nan nan nan nan \\
0.46999999999999997 0.12772894726040002 -0.11337920035092858 nan nan nan nan \\
0.47999999999999998 0.13022633372258269 -0.11594914539872508 nan nan nan nan \\
0.48999999999999999 0.13278403886639892 -0.11845289655758334 nan nan nan nan \\
0.5 0.13540268541863185 -0.12089119834426712 0.13540268541863185 -0.12089119834426712 nan nan \\
0.51000000000000001 0.13808292187169607 -0.12326462616161381 nan nan nan nan \\
0.52000000000000002 0.1408254227000158 -0.12557361567861247 nan nan nan nan \\
0.53000000000000003 0.14363088915366601 -0.12781848530104672 nan nan nan nan \\
0.54000000000000004 0.14650005006770853 -0.12999945582821187 nan nan nan nan \\
0.55000000000000004 0.14943366269016642 -0.13211666802105113 0.14943366269016642 -0.13211666802105113 nan nan \\
0.56000000000000005 0.15243251350640616 -0.13417019862507959 nan nan nan nan \\
0.56999999999999995 0.1554974189465993 -0.1361600753162594 nan nan nan nan \\
0.57999999999999996 0.15862922401464316 -0.13808629089137522 nan nan nan nan \\
0.58999999999999997 0.16182882150126557 -0.13994882057101182 nan nan nan nan \\
0.59999999999999998 0.16509710059286964 -0.14174761874167754 0.16509710059286964 -0.14174761874167754 nan nan \\
0.60999999999999999 0.16843501042756145 -0.14348266987765518 nan nan nan nan \\
0.62 0.17184350427170336 -0.14515395978054754 nan nan nan nan \\
0.63 0.17532359693520605 -0.14676153973849562 nan nan nan nan \\
0.64000000000000001 0.1788761958313683 -0.14830581340649571 nan nan nan nan \\
0.65000000000000002 0.18250336050943899 -0.14978580520072526 0.18250336050943899 -0.14978580520072526 nan nan \\
0.66000000000000003 0.18620479641417981 -0.15120303537536223 nan nan nan nan \\
0.67000000000000004 0.18998229793046306 -0.15255741774133463 nan nan nan nan \\
0.68000000000000005 0.19383698480750097 -0.15384949802253295 nan nan nan nan \\
0.68999999999999995 0.19777003133244517 -0.15508000097169755 nan nan nan nan \\
0.69999999999999996 0.20178262322356599 -0.15624987663458123 0.20178262322356599 -0.15624987663458123 nan nan \\
0.70999999999999996 0.20587593815677196 -0.15736034724453274 nan nan nan nan \\
0.71999999999999997 0.21005115157000975 -0.15841297552160888 nan nan nan nan \\
0.72999999999999998 0.21430932268713007 -0.15940970085897274 nan nan nan nan \\
0.73999999999999999 0.21865148521510031 -0.16035295238313899 nan nan nan nan \\
0.75 0.22307849555895348 -0.16124572407190571 0.22307849555895348 -0.16124572407190571 nan nan \\
0.76000000000000001 0.22759099690137799 -0.16209168812512856 nan nan nan nan \\
0.77000000000000002 0.23218931616242483 -0.16289532564761672 nan nan nan nan \\
0.78000000000000003 0.23687328878422342 -0.16366207370295038 nan nan nan nan \\
0.79000000000000004 0.24164221344954842 -0.16439849926852659 nan nan nan nan \\
0.80000000000000004 0.24649435838174 -0.16511250863714771 0.24649435838174 -0.16511250863714771 nan nan \\
0.81000000000000005 0.25142680640385578 -0.1658135329333936 nan nan nan nan \\
0.81999999999999995 0.25643490414461628 -0.1665127569556395 nan nan nan nan \\
0.82999999999999996 0.26151162840535147 -0.1672232992848014 nan nan nan nan \\
0.83999999999999997 0.26664673648961768 -0.16796032056368984 nan nan nan nan \\
0.84999999999999998 0.27182566457496765 -0.1687409701435181 0.27182566457496765 -0.1687409701435181 nan nan \\
0.85999999999999999 0.27702814982083351 -0.1695840380003083 nan nan nan nan \\
0.87 0.28222660421067702 -0.17050911905982563 nan nan nan nan \\
0.88 0.28738437606992451 -0.17153504872665634 nan nan nan nan \\
0.89000000000000001 0.29245418770517762 -0.1726774017051548 nan nan nan nan \\
0.90000000000000002 0.29737629824557804 -0.17394446339069025 0.29737629824557804 -0.17394446339069025 nan nan \\
0.91000000000000003 0.30208555867482456 -0.17533160398871131 nan nan nan nan \\
0.92000000000000004 0.30650272946260831 -0.17682297336162367 nan nan nan nan \\
0.93000000000000005 0.31055103223002634 -0.17837883455163844 nan nan nan nan \\
0.93999999999999995 0.31415940248055557 -0.17994268579895886 nan nan nan nan \\
0.94999999999999996 0.31726968973106717 -0.18144378781816756 0.31726968973106717 -0.18144378781816756 nan nan \\
0.95999999999999996 0.31984088520151732 -0.18280498853609148 nan nan nan nan \\
0.96999999999999997 0.32184915839364786 -0.18395170572473335 nan nan nan nan \\
0.97999999999999998 0.32328434938906836 -0.18481970686216853 nan nan nan nan \\
0.98999999999999999 0.32414455055741603 -0.1853604478602465 nan nan nan nan \\
1 0.32447106891313027 -0.18557496925573982 0.32447106891313027 -0.18557496925573982 0.32447106891313027 -0.18557496925573982 \\
}\FigThreeBlueData

\pgfplotstableread[row sep=\\]{
u re im mr mi er ei \\
0 0.14285714285714285 0 0.14285714285714285 0 nan nan \\
0.01 0.1428118513443988 -7.8621196760137299e-10 nan nan nan nan \\
0.02 0.14267556643804924 -9.3866479041428837e-08 nan nan nan nan \\
0.029999999999999999 0.14244735710700451 -9.209298959045508e-09 nan nan nan nan \\
0.040000000000000001 0.1421254251952335 -5.1920986341108136e-09 nan nan nan nan \\
0.050000000000000003 0.14170723616110489 -2.7235390832441262e-09 0.14170723616110489 -2.7235390832441262e-09 nan nan \\
0.059999999999999998 0.14118939635835301 -5.4675938710564877e-10 nan nan nan nan \\
0.070000000000000007 0.14056753625114338 -1.5985546178242134e-10 nan nan nan nan \\
0.080000000000000002 0.13983617831990136 9.9090884586424897e-12 nan nan nan nan \\
0.089999999999999997 0.1389884860422656 -3.4754979082237574e-11 nan nan nan nan \\
0.10000000000000001 0.13801598745145918 -1.6206277418483288e-10 0.13801598745145918 -1.6206277418483288e-10 nan nan \\
0.11 0.13690813567789759 -1.2086424693961948e-08 nan nan nan nan \\
0.12 0.13565185782641728 -3.264210919469088e-09 nan nan nan nan \\
0.13 0.13423066628926947 -1.8792164522677104e-09 nan nan nan nan \\
0.14000000000000001 0.13262336188372417 -7.7499332329880817e-08 nan nan nan nan \\
0.14999999999999999 0.13080302824414611 -4.5213788494898755e-08 0.13080302824414611 -4.5213788494898755e-08 nan nan \\
0.16 0.12873268108269836 -7.8827354356434064e-09 nan nan nan nan \\
0.17000000000000001 0.12636236040661067 -4.9954817667771584e-09 nan nan nan nan \\
0.17999999999999999 0.12361993820317289 -1.662479271719021e-09 nan nan nan nan \\
0.19 0.12039538317582586 -2.2536697968011967e-10 nan nan nan nan \\
0.20000000000000001 0.11650572034584313 -2.0800776514323877e-10 0.11650572034584313 -2.0800776514323877e-10 nan nan \\
0.20250000000000001 0.11539480025293206 -2.5839425863826289e-10 nan nan nan nan \\
0.20499999999999999 0.11421432749757569 -3.4749364146007546e-10 nan nan nan nan \\
0.20749999999999999 0.11295423205425631 -5.1640906316575341e-10 nan nan nan nan \\
0.20999999999999999 0.11160169547790369 -8.7982002524834276e-10 nan nan nan nan \\
0.21249999999999999 0.11013996716197645 -1.8075245175930724e-09 nan nan nan nan \\
0.215 0.10854643161164972 -4.5383727722568209e-09 nan nan nan nan \\
0.2175 0.10678927157293425 -1.4231709565348762e-08 nan nan nan nan \\
0.22 0.10482079104999423 1.5258277471445849e-09 nan nan nan nan \\
0.2225 0.10256487118624392 1.830549628486975e-09 nan nan nan nan \\
0.22500000000000001 0.099882224993815424 1.8539008037252001e-09 nan nan nan nan \\
0.22750000000000001 0.096461242992774074 2.1957672297830562e-09 nan nan nan nan \\
0.23000000000000001 0.0911360983909629 3.3885875468230419e-09 nan nan nan nan \\
0.23019999999999999 0.09047696574700187 3.6754730166639121e-09 nan nan nan nan \\
0.23039999999999999 0.089725527874873853 4.0921027233481257e-09 nan nan nan nan \\
0.2306 0.088829889735163439 4.7850510824429485e-09 nan nan nan nan \\
0.23080000000000001 0.087656580414829682 6.3341112685408842e-09 nan nan nan nan \\
0.23089999999999999 0.086824370833351525 8.498393044777236e-09 nan nan nan nan \\
0.23094999999999999 0.086236968579430223 1.1499859238484262e-08 nan nan nan nan \\
0.23097999999999999 0.085721603346671713 1.7177817101078491e-08 nan nan nan nan \\
0.23099 0.085469232148094584 2.3003421213288544e-08 nan nan nan nan \\
0.23100000000000001 0.085037922826563458 5.8207584331380075e-08 nan nan nan nan \\
0.23100999999999999 0.084769317169113539 0.00058958654327668537 nan nan nan nan \\
0.23102 0.084770469333374496 0.00087637117914704369 nan nan nan nan \\
0.23105000000000001 0.084773945043610729 0.0014245544163958056 nan nan nan nan \\
0.2311 0.084779746969501732 0.0020326975855266467 nan nan nan nan \\
0.23119999999999999 0.084791358905758785 0.0028875516813017918 nan nan nan nan \\
0.23139999999999999 0.084814601107313425 0.00409322524407261 nan nan nan nan \\
0.2316 0.084837864735744423 0.0050176514064101544 nan nan nan nan \\
0.23180000000000001 0.084861149195107705 0.00579697760541769 nan nan nan nan \\
0.23250000000000001 0.084942807395437123 0.0079466302231673298 nan nan nan nan \\
0.23499999999999999 0.085236502718994661 0.01300920350825813 nan nan nan nan \\
0.23749999999999999 0.085533419083134735 0.016619998202482826 nan nan nan nan \\
0.23999999999999999 0.085833575647975099 0.019598360549100864 nan nan nan nan \\
0.24249999999999999 0.086137040681185537 0.022199984409436359 nan nan nan nan \\
0.245 0.086443083891936681 0.02454513078046177 nan nan nan nan \\
0.2475 0.086752968312254736 0.026701233784320474 nan nan nan nan \\
0.25 0.087065989948319308 0.028710587604889318 nan nan nan nan \\
0.2525 0.087382238750545391 0.030601974784768522 nan nan nan nan \\
0.255 0.087701885603090501 0.032396309413249336 nan nan nan nan \\
0.25750000000000001 0.088024489334006042 0.034108415189620336 nan nan nan nan \\
0.26000000000000001 0.088350224518492476 0.035749952996675487 nan nan nan nan \\
0.27000000000000002 0.08968656595244906 0.041771692716663235 0.08968656595244906 0.041771692716663235 nan nan \\
0.28000000000000003 0.091074765081556028 0.047161140489236933 nan nan nan nan \\
0.28999999999999998 0.092515238797449342 0.052109521285500067 nan nan nan nan \\
0.29999999999999999 0.094008227244328724 0.056727275137044023 nan nan nan nan \\
0.31 0.095553972419442482 0.061084486119072946 0.095553972419442482 0.061084486119072946 nan nan \\
0.32000000000000001 0.097152740148121711 0.065228662027077597 nan nan nan nan \\
0.33000000000000002 0.098804810343348842 0.069193502450001734 nan nan nan nan \\
0.34000000000000002 0.10051047948964978 0.073003767633017982 nan nan nan nan \\
0.34999999999999998 0.10227006064824185 0.076678148254950812 0.10227006064824185 0.076678148254950812 nan nan \\
0.35999999999999999 0.10408388397640921 0.080231054930808676 nan nan nan nan \\
0.37 0.1059522973606058 0.083673786860342783 nan nan nan nan \\
0.38 0.10787566775473949 0.087015323182913676 nan nan nan nan \\
0.39000000000000001 0.10985435922847119 0.090262877216567713 nan nan nan nan \\
0.40000000000000002 0.11188882192924936 0.09342228583326817 0.11188882192924936 0.09342228583326817 nan nan \\
0.40999999999999998 0.11397944155545914 0.096498312446691922 nan nan nan nan \\
0.41999999999999998 0.11612667021910177 0.09949485251028746 nan nan nan nan \\
0.42999999999999999 0.11833097091038904 0.10241510337681048 nan nan nan nan \\
0.44 0.12059281455678095 0.1052616623214679 nan nan nan nan \\
0.45000000000000001 0.12291275771231408 0.10803678259055746 0.12291275771231408 0.10803678259055746 nan nan \\
0.46000000000000002 0.12529128039881143 0.11074214220060546 nan nan nan nan \\
0.46999999999999997 0.12772894799652132 0.1133792003353911 nan nan nan nan \\
0.47999999999999998 0.13022633400106129 0.11594914533887736 nan nan nan nan \\
0.48999999999999999 0.13278403909725187 0.11845289652594811 nan nan nan nan \\
0.5 0.13540268563294436 0.12089119833495265 0.13540268563294436 0.12089119833495265 nan nan \\
0.51000000000000001 0.13808292207707196 0.12326462617219219 nan nan nan nan \\
0.52000000000000002 0.14082542289743771 0.12557361570737388 nan nan nan nan \\
0.53000000000000003 0.14363088934195029 0.12781848534613363 nan nan nan nan \\
0.54000000000000004 0.14650005024557788 0.12999945588757103 nan nan nan nan \\
0.55000000000000004 0.14943366285808729 0.13211666809300435 0.14943366285808729 0.13211666809300435 nan nan \\
0.56000000000000005 0.15243251367094771 0.13417019871047961 nan nan nan nan \\
0.56999999999999995 0.1554974191427225 0.13616007542797107 nan nan nan nan \\
0.57999999999999996 0.15862922481942421 0.1380862908350613 nan nan nan nan \\
0.58999999999999997 0.16182881628870219 0.13994881407377099 nan nan nan nan \\
0.59999999999999998 0.16509710016033091 0.14174761809893918 0.16509710016033091 0.14174761809893918 nan nan \\
0.60999999999999999 0.16843501021933666 0.14348266929036924 nan nan nan nan \\
0.62 0.17184350441213833 0.14515396572251429 nan nan nan nan \\
0.63 0.17532359880449366 0.14676155429063162 nan nan nan nan \\
0.64000000000000001 0.17887629079786804 0.14830584894575755 nan nan nan nan \\
0.65000000000000002 0.18250336096402883 0.14978580867629832 0.18250336096402883 0.14978580867629832 nan nan \\
0.66000000000000003 0.18620479664150782 0.15120303609184721 nan nan nan nan \\
0.67000000000000004 0.18998229811842812 0.15255741827257122 nan nan nan nan \\
0.68000000000000005 0.19383698501892779 0.15384949866431971 nan nan nan nan \\
0.68999999999999995 0.19777003159332335 0.1550800019219323 nan nan nan nan \\
0.69999999999999996 0.20178262355566984 0.15624987829900427 0.20178262355566984 0.15624987829900427 nan nan \\
0.70999999999999996 0.20587593846144037 0.15736035239056623 nan nan nan nan \\
0.71999999999999997 0.21005115154959911 0.15841297545305999 nan nan nan nan \\
0.72999999999999998 0.21430932273941322 0.15940970131605758 nan nan nan nan \\
0.73999999999999999 0.21865148525407044 0.16035295404088726 nan nan nan nan \\
0.75 0.22307849539714883 0.16124572698369036 0.22307849539714883 0.16124572698369036 nan nan \\
0.76000000000000001 0.22759099638726954 0.16209169176016741 nan nan nan nan \\
0.77000000000000002 0.23218931536023729 0.1628953284654982 nan nan nan nan \\
0.78000000000000003 0.23687329167692517 0.16366206038644188 nan nan nan nan \\
0.79000000000000004 0.24164221358286156 0.1643984990657324 nan nan nan nan \\
0.80000000000000004 0.24649435812413018 0.16511250954249379 0.24649435812413018 0.16511250954249379 nan nan \\
0.81000000000000005 0.25142680569019754 0.16581353447516042 nan nan nan nan \\
0.81999999999999995 0.25643490306280353 0.16651275875729593 nan nan nan nan \\
0.82999999999999996 0.26151162694452906 0.16722330121153914 nan nan nan nan \\
0.83999999999999997 0.26664673463155125 0.16796032249013176 nan nan nan nan \\
0.84999999999999998 0.271825662282492 0.16874097196245488 0.271825662282492 0.16874097196245488 nan nan \\
0.85999999999999999 0.27702814694978301 0.16958403965045821 nan nan nan nan \\
0.87 0.28222660036569308 0.17050912045626965 nan nan nan nan \\
0.88 0.28738437028911507 0.17153504943315745 nan nan nan nan \\
0.89000000000000001 0.29245417796520634 0.17267739882090916 nan nan nan nan \\
0.90000000000000002 0.29737629347533584 0.17394446492707066 0.29737629347533584 0.17394446492707066 nan nan \\
0.91000000000000003 0.3020855728454977 0.17533158507473848 nan nan nan nan \\
0.92000000000000004 0.30650274359923463 0.1768229674430101 nan nan nan nan \\
0.93000000000000005 0.31055104641425541 0.17837883249293421 nan nan nan nan \\
0.93999999999999995 0.3141594196030425 0.17994268639272504 nan nan nan nan \\
0.94999999999999996 0.31726971523761222 0.18144379241943254 0.31726971523761222 0.18144379241943254 nan nan \\
0.95999999999999996 0.31984093252529855 0.18280500341714828 nan nan nan nan \\
0.96999999999999997 0.32184923013093325 0.1839517425680135 nan nan nan nan \\
0.97999999999999998 0.32328438188785613 0.18481973101871479 nan nan nan nan \\
0.98999999999999999 0.32414455792005614 0.18536045403422705 nan nan nan nan \\
1 0.32447076732629487 0.18557657346168599 0.32447076732629487 0.18557657346168599 0.32447076732629487 0.18557657346168599 \\
}\FigThreeOrangeData

\pgfplotstableread[row sep=\\]{
u re im mr mi er ei \\
0 0 0 0 0 nan nan \\
0.050000000000000003 -0.0018667253516652998 8.7877069782239971e-16 -0.0018667253516652998 8.7877069782239971e-16 nan nan \\
0.10000000000000001 -0.0070545809183154059 3.1615428168434965e-17 -0.0070545809183154059 3.1615428168434965e-17 nan nan \\
0.14999999999999999 -0.01462398677977351 -1.4756057429314843e-16 -0.01462398677977351 -1.4756057429314843e-16 nan nan \\
0.20000000000000001 -0.023610202536790369 4.7050979404390188e-17 -0.023610202536790369 4.7050979404390188e-17 nan nan \\
0.23100000000000001 -0.029557900654071771 -1.042596806048125e-15 nan nan nan nan \\
0.23100173140221161 -0.029558237432980194 -1.0424856925612313e-15 -0.029558237432980194 -1.0424856925612313e-15 nan nan \\
0.23100999999999999 -0.029559845781750031 -1.041955430369813e-15 nan nan nan nan \\
0.25 -0.033272927779531905 -5.0590559682453536e-16 nan nan nan nan \\
0.29999999999999999 -0.043110554126334445 2.8606486395331265e-16 -0.043110554126334445 2.8606486395331265e-16 nan nan \\
0.34999999999999998 -0.052797386369333288 3.2144276175182425e-15 -0.052797386369333288 3.2144276175182425e-15 nan nan \\
0.40000000000000002 -0.062122451512215959 5.3096399859528756e-17 -0.062122451512215959 5.3096399859528756e-17 nan nan \\
0.45000000000000001 -0.070946739679607643 -1.3367379888996047e-18 -0.070946739679607643 -1.3367379888996047e-18 nan nan \\
0.5 -0.079176190180391778 -7.9911137794925319e-17 -0.079176190180391778 -7.9911137794925319e-17 nan nan \\
0.55000000000000004 -0.086745135685658531 -1.0872273485624098e-14 -0.086745135685658531 -1.0872273485624098e-14 nan nan \\
0.59999999999999998 -0.093606200461861369 2.2975476478970256e-13 -0.093606200461861369 2.2975476478970256e-13 nan nan \\
0.65000000000000002 -0.099724112787404992 1.1770969972208916e-16 -0.099724112787404992 1.1770969972208916e-16 nan nan \\
0.69999999999999996 -0.10507190027764132 -2.353928231942417e-17 -0.10507190027764132 -2.353928231942417e-17 nan nan \\
0.75 -0.1096285515088572 -3.4512989229061001e-16 -0.1096285515088572 -3.4512989229061001e-16 nan nan \\
0.80000000000000004 -0.11337758918032251 -1.025964864882617e-15 -0.11337758918032251 -1.025964864882617e-15 nan nan \\
0.81999999999999995 -0.11464809467093014 -1.6514428484730583e-15 nan nan nan nan \\
0.82499999999999996 -0.11494512467168325 -1.9210557946537375e-15 nan nan nan nan \\
0.82999999999999996 -0.11523389618692134 -2.2970260583474481e-15 nan nan nan nan \\
0.84999999999999998 -0.11630621295710097 -1.4034270307940021e-14 -0.11630621295710097 -1.4034270307940021e-14 nan nan \\
0.90000000000000002 -0.11840479742571111 -2.6526344209032118e-16 -0.11840479742571111 -2.6526344209032118e-16 nan nan \\
0.94999999999999996 -0.11966660880659355 6.8140114708734388e-16 -0.11966660880659355 6.8140114708734388e-16 nan nan \\
1 -0.12008748183695754 8.0651064812614332e-09 -0.12008748183695754 8.0651064812614332e-09 -0.12008748183695754 8.0651064812614332e-09 \\
}\FigThreeGreenData

\pgfplotstableread[row sep=\\]{
u re im mr mi er ei \\
0 0.14285714285714285 0 0.14285714285714285 0 nan nan \\
0.050000000000000003 0.14576538483646817 2.7942644351931551e-15 0.14576538483646817 2.7942644351931551e-15 nan nan \\
0.10000000000000001 0.15407761761939864 4.6573429091144791e-16 0.15407761761939864 4.6573429091144791e-16 nan nan \\
0.14999999999999999 0.16685340171327084 -1.3627443380727145e-15 0.16685340171327084 -1.3627443380727145e-15 nan nan \\
0.20000000000000001 0.18312537168137391 1.3085272761842854e-15 0.18312537168137391 1.3085272761842854e-15 nan nan \\
0.23100000000000001 0.19463194536125564 -3.7273096640462588e-14 nan nan nan nan \\
0.23100173140221161 0.19463261482599292 -3.7268017056835111e-14 0.19463261482599292 -3.7268017056835111e-14 nan nan \\
0.23100999999999999 0.19463581200516988 -3.7243781439247113e-14 nan nan nan nan \\
0.25 0.20214684836048954 -1.8180949628577569e-14 nan nan nan nan \\
0.29999999999999999 0.22340286562954623 2.3803272859967579e-14 0.22340286562954623 2.3803272859967579e-14 nan nan \\
0.34999999999999998 0.24654238066301767 4.9500112106532402e-13 0.24654238066301767 4.9500112106532402e-13 nan nan \\
0.40000000000000002 0.27130941074802445 7.3769026346760518e-15 0.27130941074802445 7.3769026346760518e-15 nan nan \\
0.45000000000000001 0.29748859168081537 2.6301207444324769e-16 0.29748859168081537 2.6301207444324769e-16 nan nan \\
0.5 0.32486049778514453 -1.4648451332364099e-13 0.32486049778514453 -1.4648451332364099e-13 nan nan \\
0.55000000000000004 0.35315884136560433 -4.0423754796012152e-10 0.35315884136560433 -4.0423754796012152e-10 nan nan \\
0.59999999999999998 0.38202293504578488 4.3713387207119974e-09 0.38202293504578488 4.3713387207119974e-09 nan nan \\
0.65000000000000002 0.4109427339002088 3.9059464791884453e-11 0.4109427339002088 3.9059464791884453e-11 nan nan \\
0.69999999999999996 0.43920277349037723 2.7563477672019802e-11 0.43920277349037723 2.7563477672019802e-11 nan nan \\
0.75 0.46585017976913662 -6.7744915173355489e-11 0.46585017976913662 -6.7744915173355489e-11 nan nan \\
0.80000000000000004 0.48973253889471091 -1.149236203486117e-09 0.48973253889471091 -1.149236203486117e-09 nan nan \\
0.81999999999999995 0.49823800964838461 -3.2315648442900734e-09 nan nan nan nan \\
0.82499999999999996 0.50025616172870346 -4.3574223150792308e-09 nan nan nan nan \\
0.82999999999999996 0.50222883642737792 -6.1079355977163795e-09 nan nan nan nan \\
0.84999999999999998 0.5096440731252434 7.149238111980681e-08 0.5096440731252434 7.149238111980681e-08 nan nan \\
0.90000000000000002 0.52454509850319242 -1.9905375140932341e-09 0.52454509850319242 -1.9905375140932341e-09 nan nan \\
0.94999999999999996 0.53372578152018479 2.1581959545648071e-08 0.53372578152018479 2.1581959545648071e-08 nan nan \\
}\FigThreeCyanData

\pgfplotstableread[row sep=\\]{
u re im mr mi er ei \\
0 0.42857142857142855 0 0.42857142857142855 0 nan nan \\
0.01 0.42745789067908518 -5.1617992391736317e-08 nan nan nan nan \\
0.02 0.42424280710462237 -1.4132144647185319e-06 nan nan nan nan \\
0.029999999999999999 0.41921903406474353 -1.2553520187511743e-05 nan nan nan nan \\
0.040000000000000001 0.41281814992409155 -1.9224825886341737e-06 nan nan nan nan \\
0.050000000000000003 0.40538516071035152 -5.2840367386896705e-07 0.40538516071035152 -5.2840367386896705e-07 nan nan \\
0.059999999999999998 0.39722033343012275 -7.5682066020648033e-08 nan nan nan nan \\
0.070000000000000007 0.38855153606381482 -2.2630302710758573e-08 nan nan nan nan \\
0.080000000000000002 0.37954728250572867 1.5660301451291026e-09 nan nan nan nan \\
0.089999999999999997 0.3703272618493389 3.9205618665935476e-10 nan nan nan nan \\
0.10000000000000001 0.36097741679264217 -7.1261206713543483e-09 0.36097741679264217 -7.1261206713543483e-09 nan nan \\
0.11 0.35155818769537978 -6.40770699469324e-07 nan nan nan nan \\
0.12 0.34211621727270725 -2.2658087881797907e-07 nan nan nan nan \\
0.13 0.33268567976297658 -1.4628274666039291e-07 nan nan nan nan \\
0.14000000000000001 0.32329571986472322 -9.4736707063413164e-07 nan nan nan nan \\
0.14999999999999999 0.3139518126651471 -7.2108087179085458e-07 0.3139518126651471 -7.2108087179085458e-07 nan nan \\
0.16 0.30467810444691912 -1.567919640769405e-07 nan nan nan nan \\
0.17000000000000001 0.29548388917606383 -8.0812523662012325e-08 nan nan nan nan \\
0.17999999999999999 0.28637671518092783 -2.0795725974595525e-08 nan nan nan nan \\
0.19 0.27736248925075596 -2.4081926117251277e-09 nan nan nan nan \\
0.20000000000000001 0.26844631649502199 -2.0209475771432789e-09 0.26844631649502199 -2.0209475771432789e-09 nan nan \\
0.20250000000000001 0.26623308375249927 -2.4564845758463158e-09 nan nan nan nan \\
0.20499999999999999 0.26402628398144301 -3.2365827922625667e-09 nan nan nan nan \\
0.20749999999999999 0.26182596690534499 -4.7266925501525374e-09 nan nan nan nan \\
0.20999999999999999 0.25963218205080268 -7.9749544944405247e-09 nan nan nan nan \\
0.21249999999999999 0.25744498655843701 -1.6577137177331182e-08 nan nan nan nan \\
0.215 0.25526450004160461 -4.5706822834721049e-08 nan nan nan nan \\
0.2175 0.2530942426706933 -1.1959118711136407e-06 nan nan nan nan \\
0.22 0.25092253258389952 1.081083074229172e-08 nan nan nan nan \\
0.2225 0.24876215816503697 8.5381695472020371e-09 nan nan nan nan \\
0.22500000000000001 0.24660836562049154 8.4337304346185092e-09 nan nan nan nan \\
0.22750000000000001 0.24446130820000944 8.9291443570222015e-09 nan nan nan nan \\
0.23000000000000001 0.24232103995234791 9.2989338871405392e-09 nan nan nan nan \\
0.23019999999999999 0.24215011297023001 9.3133622597613294e-09 nan nan nan nan \\
0.23039999999999999 0.2419792297045571 9.3250608095978779e-09 nan nan nan nan \\
0.2306 0.2418083901736891 9.3339309850381545e-09 nan nan nan nan \\
0.23080000000000001 0.24163759439598231 9.3398696044305955e-09 nan nan nan nan \\
0.23089999999999999 0.24155221292030454 9.3417056989801763e-09 nan nan nan nan \\
0.23094999999999999 0.24150952628676511 9.3423342452085497e-09 nan nan nan nan \\
0.23097999999999999 0.24148391562018298 9.3426176784482329e-09 nan nan nan nan \\
0.23099 0.24147537895025967 9.3426964510629731e-09 nan nan nan nan \\
0.23100000000000001 0.24146684238980887 9.3427673460508089e-09 nan nan nan nan \\
0.23100173140221161 0.24146536437896257 9.342778819670122e-09 0.24146536437896257 9.342778819670122e-09 nan nan \\
0.23100999999999999 0.24145830593883294 9.3428303482898001e-09 nan nan nan nan \\
0.23102 0.24144976959733414 9.3428854426004215e-09 nan nan nan nan \\
0.23105000000000001 0.24142416122972365 9.3430031252927833e-09 nan nan nan nan \\
0.2311 0.24138148280679753 9.3430396707295133e-09 nan nan nan nan \\
0.23119999999999999 0.24129613417357601 9.3425072221703394e-09 nan nan nan nan \\
0.23139999999999999 0.24112546976574525 9.3389630249711747e-09 nan nan nan nan \\
0.2316 0.24095484918485197 9.3319977309901096e-09 nan nan nan nan \\
0.23180000000000001 0.24078427244952466 9.3214618496615148e-09 nan nan nan nan \\
0.23250000000000001 0.24018759943156145 9.2539551395359579e-09 nan nan nan nan \\
0.23499999999999999 0.238061023583063 8.4651576614423975e-09 nan nan nan nan \\
0.23749999999999999 0.23594135524236731 5.8949178929265291e-09 nan nan nan nan \\
0.23999999999999999 0.23382865570428368 -2.1441708864101996e-09 nan nan nan nan \\
0.24249999999999999 0.23172303630843893 -3.070049677636296e-08 nan nan nan nan \\
0.245 0.22962473381770551 -1.5361068849931435e-07 nan nan nan nan \\
0.2475 0.22753400326866774 -6.9386307192981773e-07 nan nan nan nan \\
0.25 0.22544911524896316 -1.2263059737732017e-06 nan nan nan nan \\
0.2525 0.22336915015232253 -4.5748413462169651e-07 nan nan nan nan \\
0.255 0.22129678071115053 -1.2348926124135714e-07 nan nan nan nan \\
0.25750000000000001 0.2192319532765678 -4.3991893713374493e-08 nan nan nan nan \\
0.26000000000000001 0.21717901610696441 1.3715643782059608e-06 nan nan nan nan \\
0.27000000000000002 0.20901640359740753 5.437726625977913e-11 0.20901640359740753 5.437726625977913e-11 nan nan \\
0.28000000000000003 0.20097100734776668 3.5645701000382876e-11 nan nan nan nan \\
0.28999999999999998 0.19303892394716404 -7.8188733541687824e-08 nan nan nan nan \\
0.29999999999999999 0.18522115393329455 7.2494240209095172e-11 nan nan nan nan \\
0.31 0.17751724215974557 -5.3379059420706018e-11 0.17751724215974557 -5.3379059420706018e-11 nan nan \\
0.32000000000000001 0.16992756624511865 -1.9571739640683068e-10 nan nan nan nan \\
0.33000000000000002 0.16245220737440316 -1.0357784976034696e-10 nan nan nan nan \\
0.34000000000000002 0.15509118790112547 -4.979764198953108e-11 nan nan nan nan \\
0.34999999999999998 0.14784449247288897 -3.4152515847878194e-11 0.14784449247288897 -3.4152515847878194e-11 nan nan \\
0.35999999999999999 0.14071207071124978 -3.9310968309763993e-11 nan nan nan nan \\
0.37 0.13369384384518293 -6.7214192089784145e-11 nan nan nan nan \\
0.38 0.12678970908546833 -1.5284604144240947e-11 nan nan nan nan \\
0.39000000000000001 0.11999958686406943 1.4377365427922001e-11 nan nan nan nan \\
0.40000000000000002 0.1133232585822167 6.484956527830036e-12 0.1133232585822167 6.484956527830036e-12 nan nan \\
0.40999999999999998 0.10676065319501188 1.4866250374072486e-11 nan nan nan nan \\
0.41999999999999998 0.10031160537143143 1.5234638742068449e-11 nan nan nan nan \\
0.42999999999999999 0.093975961594163271 1.091449655680814e-11 nan nan nan nan \\
0.44 0.087753583188141249 6.5785334731319866e-10 nan nan nan nan \\
0.45000000000000001 0.081644247483218343 7.4556935226403748e-11 0.081644247483218343 7.4556935226403748e-11 nan nan \\
0.46000000000000002 0.075647857057498505 3.5173984875972719e-11 nan nan nan nan \\
0.46999999999999997 0.069764235550399567 1.0694533683357845e-10 nan nan nan nan \\
0.47999999999999998 0.063993218781548275 3.5217745391523537e-11 nan nan nan nan \\
0.48999999999999999 0.058334653504274782 2.3101620703746525e-11 nan nan nan nan \\
0.5 0.052788383724300107 1.6812000078306816e-11 0.052788383724300107 1.6812000078306816e-11 nan nan \\
0.51000000000000001 0.047354256950450133 1.2651524659300506e-11 nan nan nan nan \\
0.52000000000000002 0.042032123678697665 9.5824517540622735e-12 nan nan nan nan \\
0.53000000000000003 0.036821837612535657 7.2076355914038701e-12 nan nan nan nan \\
0.54000000000000004 0.031723255857424991 5.3406179082563925e-12 nan nan nan nan \\
0.55000000000000004 0.026736239072098846 3.8757877203082706e-12 0.026736239072098846 3.8757877203082706e-12 nan nan \\
0.56000000000000005 0.021860651579593594 2.7442195821709842e-12 nan nan nan nan \\
0.56999999999999995 0.017096361439835478 1.7816473428745393e-12 nan nan nan nan \\
0.57999999999999996 0.012443240421965632 -1.0923169475686262e-11 nan nan nan nan \\
0.58999999999999997 0.0079011647079525286 -1.4800883522928006e-10 nan nan nan nan \\
0.59999999999999998 0.0034700131517724315 -8.9954346429228002e-12 0.0034700131517724315 -8.9954346429228002e-12 nan nan \\
0.60999999999999999 -0.00085033073838051759 -4.903161947841621e-12 nan nan nan nan \\
0.62 -0.0050599799449506716 3.1794415617381253e-11 nan nan nan nan \\
0.63 -0.0091590434940528875 6.7298169329631383e-11 nan nan nan nan \\
0.64000000000000001 -0.013147633353743457 1.552153722162961e-09 nan nan nan nan \\
0.65000000000000002 -0.017025826595024 9.5264999160371117e-12 -0.017025826595024 9.5264999160371117e-12 nan nan \\
0.66000000000000003 -0.020793749483003904 1.8889671429725273e-12 nan nan nan nan \\
0.67000000000000004 -0.02445148556349315 1.2213770064507338e-12 nan nan nan nan \\
0.68000000000000005 -0.027999126038792274 1.223345584173569e-12 nan nan nan nan \\
0.68999999999999995 -0.031436758391850089 1.4370718454142098e-12 nan nan nan nan \\
0.69999999999999996 -0.034764466667146963 1.9447256835995311e-12 -0.034764466667146963 1.9447256835995311e-12 nan nan \\
0.70999999999999996 -0.037982331560546562 4.661162726188782e-12 nan nan nan nan \\
0.71999999999999997 -0.041090430429281614 -3.5195552649941702e-14 nan nan nan nan \\
0.72999999999999998 -0.044088837540470882 1.5008172345010972e-13 nan nan nan nan \\
0.73999999999999999 -0.04697762384550061 3.9680519955376601e-13 nan nan nan nan \\
0.75 -0.049756857260970545 4.502158168591266e-13 -0.049756857260970545 4.502158168591266e-13 nan nan \\
0.76000000000000001 -0.052426602642255134 3.0899289302115582e-13 nan nan nan nan \\
0.77000000000000002 -0.054986921849503753 1.881538954146317e-13 nan nan nan nan \\
0.78000000000000003 -0.057437873789839597 3.0576870882531043e-12 nan nan nan nan \\
0.79000000000000004 -0.059779514513792972 2.3122975455242371e-13 nan nan nan nan \\
0.80000000000000004 -0.062011897194110946 3.2935418847636398e-13 -0.062011897194110946 3.2935418847636398e-13 nan nan \\
0.81000000000000005 -0.064135072229222559 1.1099044912598442e-13 nan nan nan nan \\
0.81999999999999995 -0.066149087269039897 -7.5640628968667165e-14 nan nan nan nan \\
0.82999999999999996 -0.068053987259887594 -2.231063215215429e-13 nan nan nan nan \\
0.83999999999999997 -0.069849814485806475 -3.3631818332517046e-13 nan nan nan nan \\
0.84999999999999998 -0.071536608607717961 -4.1551304130573284e-13 -0.071536608607717961 -4.1551304130573284e-13 nan nan \\
0.85999999999999999 -0.073114406700334464 -4.7205023683257513e-13 nan nan nan nan \\
0.87 -0.074583243286904882 -5.3062329106628187e-13 nan nan nan nan \\
0.88 -0.075943150372130536 -6.2284819198660813e-13 nan nan nan nan \\
0.89000000000000001 -0.07719415747501486 -7.7429996232848848e-13 nan nan nan nan \\
0.90000000000000002 -0.07833629167826521 -5.7663306476883413e-13 -0.07833629167826521 -5.7663306476883413e-13 nan nan \\
0.91000000000000003 -0.079369575646612789 6.1350457615616131e-10 nan nan nan nan \\
0.92000000000000004 -0.08029403695195507 5.7843930938749565e-12 nan nan nan nan \\
0.93000000000000005 -0.081109690370117768 3.2774119319657463e-12 nan nan nan nan \\
0.93999999999999995 -0.081816554867972333 3.1991680526916975e-12 nan nan nan nan \\
0.94999999999999996 -0.082414645546370915 4.4061130040631138e-12 -0.082414645546370915 4.4061130040631138e-12 nan nan \\
0.95999999999999996 -0.082903975144218572 7.957234251974067e-12 nan nan nan nan \\
0.96999999999999997 -0.083284554069873409 1.061926557825702e-11 nan nan nan nan \\
0.97999999999999998 -0.083556390394093863 4.4963098123726248e-12 nan nan nan nan \\
0.98999999999999999 -0.08371948988071945 1.0507358533082572e-12 nan nan nan nan \\
1 -0.083773357306419466 2.1278676502616041e-08 -0.083773357306419466 2.1278676502616041e-08 -0.083773357306419466 2.1278676502616041e-08 \\
}\FigThreePurpleData

\pgfplotstableread[row sep=\\]{
u re im mr mi er ei \\
0.84999999999999998 0.56311110977579837 -5.159851357460251e-07 0.56311110977579837 -5.159851357460251e-07 nan nan \\
0.85999999999999999 0.54804138651306067 -4.4802841594007183e-07 nan nan nan nan \\
0.87 0.53334836095039539 -4.4259267921157456e-07 nan nan nan nan \\
0.88 0.51908753992595169 -5.5287979200999245e-07 nan nan nan nan \\
0.89000000000000001 0.5053137091559402 -1.2562626744122851e-06 nan nan nan nan \\
0.90000000000000002 0.49245368155786379 1.0787419539213575e-06 0.49245368155786379 1.0787419539213575e-06 nan nan \\
0.91000000000000003 0.47999395138044865 4.0499630758975207e-07 nan nan nan nan \\
0.92000000000000004 0.46854308715084003 2.5708895938080581e-07 nan nan nan nan \\
0.93000000000000005 0.45815667617808786 2.1687293296087791e-07 nan nan nan nan \\
0.93999999999999995 0.44896568394904696 2.299594360754409e-07 nan nan nan nan \\
0.94999999999999996 0.44108107635006327 3.105588371453397e-07 0.44108107635006327 3.105588371453397e-07 nan nan \\
0.95999999999999996 0.4345812517181224 5.4046424505715615e-07 nan nan nan nan \\
0.96999999999999997 0.42951089883461241 8.3996841431111424e-07 nan nan nan nan \\
0.97999999999999998 0.42588894006210293 4.185693990653927e-07 nan nan nan nan \\
0.98999999999999999 0.42371841145334588 9.655523365256324e-08 nan nan nan nan \\
1 0.42298868450097094 -1.0866298653931237e-12 0.42298868450097094 -1.0866298653931237e-12 0.42298868450097094 -1.0866298653931237e-12 \\
}\FigThreeGoldData

\pgfplotstableread[row sep=\\]{
u re im mr mi er ei \\
0 0 0 nan nan nan nan \\
0.01 0 0 nan nan nan nan \\
0.02 0 0 nan nan nan nan \\
0.029999999999999999 0 0 nan nan nan nan \\
0.040000000000000001 0 0 nan nan nan nan \\
0.050000000000000003 0 0 nan nan nan nan \\
0.059999999999999998 0 0 nan nan nan nan \\
0.070000000000000007 0 0 nan nan nan nan \\
0.080000000000000002 0 0 nan nan nan nan \\
0.089999999999999997 0 0 nan nan nan nan \\
0.10000000000000001 0 0 nan nan nan nan \\
0.11 0 0 nan nan nan nan \\
0.12 0 0 nan nan nan nan \\
0.13 0 0 nan nan nan nan \\
0.14000000000000001 0 0 nan nan nan nan \\
0.14999999999999999 0 0 nan nan nan nan \\
0.16 0 0 nan nan nan nan \\
0.17000000000000001 0 0 nan nan nan nan \\
0.17999999999999999 0 0 nan nan nan nan \\
0.19 0 0 nan nan nan nan \\
0.20000000000000001 0 0 nan nan nan nan \\
0.20250000000000001 0 0 nan nan nan nan \\
0.20499999999999999 0 0 nan nan nan nan \\
0.20749999999999999 0 0 nan nan nan nan \\
0.20999999999999999 0 0 nan nan nan nan \\
0.21249999999999999 0 0 nan nan nan nan \\
0.215 0 0 nan nan nan nan \\
0.2175 0 0 nan nan nan nan \\
0.22 0 0 nan nan nan nan \\
0.2225 0 0 nan nan nan nan \\
0.22500000000000001 0 0 nan nan nan nan \\
0.22750000000000001 0 0 nan nan nan nan \\
0.23000000000000001 0 0 nan nan nan nan \\
0.23019999999999999 0 0 nan nan nan nan \\
0.23039999999999999 0 0 nan nan nan nan \\
0.2306 0 0 nan nan nan nan \\
0.23080000000000001 0 0 nan nan nan nan \\
0.23089999999999999 0 0 nan nan nan nan \\
0.23094999999999999 0 0 nan nan nan nan \\
0.23097999999999999 0 0 nan nan nan nan \\
0.23099 0 0 nan nan nan nan \\
0.23100000000000001 0 0 nan nan nan nan \\
0.23100173140221161 0 0 0 0 nan nan \\
0.23100999999999999 0 0 nan nan nan nan \\
0.23102 0 0 nan nan nan nan \\
0.23105000000000001 0 0 nan nan nan nan \\
0.2311 0 0 nan nan nan nan \\
0.23119999999999999 0 0 nan nan nan nan \\
0.23139999999999999 0 0 nan nan nan nan \\
0.2316 0 0 nan nan nan nan \\
0.23180000000000001 0 0 nan nan nan nan \\
0.23250000000000001 0 0 nan nan nan nan \\
0.23499999999999999 0 0 nan nan nan nan \\
0.23749999999999999 0 0 nan nan nan nan \\
0.23999999999999999 0 0 nan nan nan nan \\
0.24249999999999999 0 0 nan nan nan nan \\
0.245 0 0 nan nan nan nan \\
0.2475 0 0 nan nan nan nan \\
0.25 0 0 nan nan nan nan \\
0.2525 0 0 nan nan nan nan \\
0.255 0 0 nan nan nan nan \\
0.25750000000000001 0 0 nan nan nan nan \\
0.26000000000000001 0 0 nan nan nan nan \\
0.27000000000000002 0 0 nan nan nan nan \\
0.28000000000000003 0 0 nan nan nan nan \\
0.28999999999999998 0 0 nan nan nan nan \\
0.29999999999999999 0 0 nan nan nan nan \\
0.31 0 0 nan nan nan nan \\
0.32000000000000001 0 0 nan nan nan nan \\
0.33000000000000002 0 0 nan nan nan nan \\
0.34000000000000002 0 0 nan nan nan nan \\
0.34999999999999998 0 0 nan nan nan nan \\
0.35999999999999999 0 0 nan nan nan nan \\
0.37 0 0 nan nan nan nan \\
0.38 0 0 nan nan nan nan \\
0.39000000000000001 0 0 nan nan nan nan \\
0.40000000000000002 0 0 nan nan nan nan \\
0.40999999999999998 0 0 nan nan nan nan \\
0.41999999999999998 0 0 nan nan nan nan \\
0.42999999999999999 0 0 nan nan nan nan \\
0.44 0 0 nan nan nan nan \\
0.45000000000000001 0 0 nan nan nan nan \\
0.46000000000000002 0 0 nan nan nan nan \\
0.46999999999999997 0 0 nan nan nan nan \\
0.47999999999999998 0 0 nan nan nan nan \\
0.48999999999999999 0 0 nan nan nan nan \\
0.5 0 0 nan nan nan nan \\
0.51000000000000001 0 0 nan nan nan nan \\
0.52000000000000002 0 0 nan nan nan nan \\
0.53000000000000003 0 0 nan nan nan nan \\
0.54000000000000004 0 0 nan nan nan nan \\
0.55000000000000004 0 0 nan nan nan nan \\
0.56000000000000005 0 0 nan nan nan nan \\
0.56999999999999995 0 0 nan nan nan nan \\
0.57999999999999996 0 0 nan nan nan nan \\
0.58999999999999997 0 0 nan nan nan nan \\
0.59999999999999998 0 0 nan nan nan nan \\
0.60999999999999999 0 0 nan nan nan nan \\
0.62 0 0 nan nan nan nan \\
0.63 0 0 nan nan nan nan \\
0.64000000000000001 0 0 nan nan nan nan \\
0.65000000000000002 0 0 nan nan nan nan \\
0.66000000000000003 0 0 nan nan nan nan \\
0.67000000000000004 0 0 nan nan nan nan \\
0.68000000000000005 0 0 nan nan nan nan \\
0.68999999999999995 0 0 nan nan nan nan \\
0.69999999999999996 0 0 nan nan nan nan \\
0.70999999999999996 0 0 nan nan nan nan \\
0.71999999999999997 0 0 nan nan nan nan \\
0.72999999999999998 0 0 nan nan nan nan \\
0.73999999999999999 0 0 nan nan nan nan \\
0.75 0 0 nan nan nan nan \\
0.76000000000000001 0 0 nan nan nan nan \\
0.77000000000000002 0 0 nan nan nan nan \\
0.78000000000000003 0 0 nan nan nan nan \\
0.79000000000000004 0 0 nan nan nan nan \\
0.80000000000000004 0 0 nan nan nan nan \\
0.81000000000000005 0 0 nan nan nan nan \\
0.81999999999999995 0 0 nan nan nan nan \\
0.82499999999999996 0 0 nan nan nan nan \\
0.82999999999999996 0 0 nan nan nan nan \\
0.83999999999999997 0 0 nan nan nan nan \\
0.84999999999999998 0 0 nan nan nan nan \\
0.85999999999999999 0 0 nan nan nan nan \\
0.87 0 0 nan nan nan nan \\
0.88 0 0 nan nan nan nan \\
0.89000000000000001 0 0 nan nan nan nan \\
0.90000000000000002 0 0 nan nan nan nan \\
0.91000000000000003 0 0 nan nan nan nan \\
0.92000000000000004 0 0 nan nan nan nan \\
0.93000000000000005 0 0 nan nan nan nan \\
0.93999999999999995 0 0 nan nan nan nan \\
0.94999999999999996 0 0 nan nan nan nan \\
0.95999999999999996 0 0 nan nan nan nan \\
0.96999999999999997 0 0 nan nan nan nan \\
0.97999999999999998 0 0 nan nan nan nan \\
0.98999999999999999 0 0 nan nan nan nan \\
1 0 0 nan nan nan nan \\
}\FigThreeIdentityData

\begin{tikzpicture}[x=1in,y=1in]
\path[use as bounding box] (0,0) rectangle
  (\FigThreeCanvasWidth,\FigThreeCanvasHeight);
\pgfplotsset{
  fig three axes/.style={
    scale only axis,width=\FigThreeAxesWidth in,
    xmin=-.012,xmax=1.012,
    axis x line*=bottom,axis y line*=left,
    axis line style={draw=black,line width=.65bp},
    tick align=outside,major tick length=3bp,
    tick style={draw=black,line width=.65bp},
    tick label style={font=\FigThreeTickFont},
    scaled ticks=false,xtick={0,.4,.6,.8,1},
    xticklabels={$0$,$0.4$,$0.6$,$0.8$,$1.0$},
    xticklabel style={yshift=1bp},yticklabel style={xshift=1bp},
    axis background/.style={fill=white},clip=true,
    unbounded coords=jump,filter discard warning=false,
  },
  fig three circles/.style={
    only marks,mark=*,mark size=\FigThreeCircleRadius,
    mark options={solid,fill=white,line width=.65bp},
  },
  fig three diamonds/.style={
    only marks,mark=figthreediamond,mark size=\FigThreeDiamondRadius,
    mark options={solid,draw=white,line width=.4bp},
  },
  fig three star/.style={
    only marks,mark=figthreefilledstar,mark size=\FigThreeStarRadius,
    mark options={solid,draw=white,fill=FigThreeEPColor,line width=.45bp},
  },
}

\begin{axis}[
  fig three axes,
  at={(\FigThreeAxesLeft in,\FigThreeRealBottom in)},anchor=south west,
  height=\FigThreeRealHeight in,ymin=-.14,ymax=.555,
  ytick={-.1,0,.1,.2,.3,.4,.5},
  yticklabels={$-0.1$,$0.0$,$0.1$,$0.2$,$0.3$,$0.4$,$0.5$},
  xticklabels=\empty,
]
\fill[FigThreeBand] (axis cs:-.012,.5) rectangle (axis cs:1.012,.565);
\draw[FigThreeGuide,line width=.7bp,dash pattern=on 1.4bp off 2.1bp]
  (axis cs:\FigThreeEP,-.14)--(axis cs:\FigThreeEP,.555);
\draw[FigThreeCutoff,line width=.65bp,dash pattern=on 2.6bp off 1.95bp]
  (axis cs:-.012,.5)--(axis cs:1.012,.5);
\addplot[FigThreeBlue,line width=\FigThreeCurveWidth,no marks] table[x=u,y=re]{\FigThreeBlueData};
\addplot[fig three circles,draw=FigThreeBlue] table[x=u,y=mr]{\FigThreeBlueData};
\addplot[fig three diamonds,fill=FigThreeBlue] table[x=u,y=er]{\FigThreeBlueData};
\addplot[FigThreeOrange,line width=\FigThreeCurveWidth,no marks] table[x=u,y=re]{\FigThreeOrangeData};
\addplot[fig three circles,draw=FigThreeOrange] table[x=u,y=mr]{\FigThreeOrangeData};
\addplot[fig three diamonds,fill=FigThreeOrange] table[x=u,y=er]{\FigThreeOrangeData};
\addplot[FigThreeGreen,line width=\FigThreeCurveWidth,no marks] table[x=u,y=re]{\FigThreeGreenData};
\addplot[fig three circles,draw=FigThreeGreen] table[x=u,y=mr]{\FigThreeGreenData};
\addplot[fig three diamonds,fill=FigThreeGreen] table[x=u,y=er]{\FigThreeGreenData};
\addplot[FigThreeCyan,line width=\FigThreeCurveWidth,no marks] table[x=u,y=re]{\FigThreeCyanData};
\addplot[fig three circles,draw=FigThreeCyan] table[x=u,y=mr]{\FigThreeCyanData};
\addplot[FigThreePurple,line width=\FigThreeCurveWidth,no marks] table[x=u,y=re]{\FigThreePurpleData};
\addplot[fig three circles,draw=FigThreePurple] table[x=u,y=mr]{\FigThreePurpleData};
\addplot[fig three diamonds,fill=FigThreePurple] table[x=u,y=er]{\FigThreePurpleData};
\addplot[FigThreeGold,line width=\FigThreeCurveWidth,no marks] table[x=u,y=re]{\FigThreeGoldData};
\addplot[fig three circles,draw=FigThreeGold] table[x=u,y=mr]{\FigThreeGoldData};
\addplot[fig three diamonds,fill=FigThreeGold] table[x=u,y=er]{\FigThreeGoldData};
\addplot[FigThreeIdentity,line width=\FigThreeCurveWidth,no marks,dash pattern=on 4.44bp off 1.92bp] table[x=u,y=re]{\FigThreeIdentityData};
\addplot[fig three circles,draw=FigThreeIdentity] table[x=u,y=mr]{\FigThreeIdentityData};
\draw[FigThreeLeader,line width=.65bp]
  (axis cs:\FigThreeEP,\FigThreeEPH)
  --(axis cs:\FigThreeEPBendX,\FigThreeEPLabelY)
  --(axis cs:\FigThreeEPLeaderEndX,\FigThreeEPLabelY);
\node[anchor=west,inner sep=0pt,text=FigThreeEPText,
      font=\fontsize{9bp}{10.8bp}\selectfont\bfseries]
  at (axis cs:\FigThreeEPLabelX,\FigThreeEPLabelY) {EP};
\addplot[fig three star] coordinates {(\FigThreeEP,\FigThreeEPH)};
\end{axis}

\begin{axis}[
  fig three axes,
  at={(\FigThreeAxesLeft in,\FigThreeImagBottom in)},anchor=south west,
  height=\FigThreeImagHeight in,ymin=-.205,ymax=.205,
  ytick={-.2,0,.2},yticklabels={$-0.2$,$0.0$,$0.2$},
  minor xtick={\FigThreeEP},minor tick length=3bp,
  minor tick style={draw=FigThreeEPColor,line width=.65bp},
]
\draw[black!18,line width=.6bp] (axis cs:-.012,0)--(axis cs:1.012,0);
\draw[FigThreeGuide,line width=.7bp,dash pattern=on 1.4bp off 2.1bp]
  (axis cs:\FigThreeEP,-.205)--(axis cs:\FigThreeEP,.205);
\addplot[FigThreeBlue,line width=\FigThreeCurveWidth,no marks] table[x=u,y=im]{\FigThreeBlueData};
\addplot[fig three circles,draw=FigThreeBlue] table[x=u,y=mi]{\FigThreeBlueData};
\addplot[fig three diamonds,fill=FigThreeBlue] table[x=u,y=ei]{\FigThreeBlueData};
\addplot[FigThreeOrange,line width=\FigThreeCurveWidth,no marks] table[x=u,y=im]{\FigThreeOrangeData};
\addplot[fig three circles,draw=FigThreeOrange] table[x=u,y=mi]{\FigThreeOrangeData};
\addplot[fig three diamonds,fill=FigThreeOrange] table[x=u,y=ei]{\FigThreeOrangeData};
\addplot[fig three star] coordinates {(\FigThreeEP,0)};
\end{axis}

\node[rotate=90,inner sep=0pt,font=\FigThreeLabelFont]
  at (.1318,1.971) {$\mathrm{Re}\,h$};
\node[rotate=90,inner sep=0pt,font=\FigThreeLabelFont]
  at (.1318,.783) {$\mathrm{Im}\,h$};
\node[anchor=base west,inner sep=0pt,font=\fontsize{9bp}{10.8bp}\selectfont]
  at (.15424,2.72466) {(a)};
\node[anchor=base west,inner sep=0pt,font=\fontsize{9bp}{10.8bp}\selectfont]
  at (.15424,1.0899) {(b)};
\pgfmathsetmacro{\FigThreeGuideX}
  {\FigThreeAxesLeft+(\FigThreeEP+.012)/1.024*\FigThreeAxesWidth}
\node[anchor=north,inner sep=0pt,align=center,text=FigThreeEPText,
      font=\fontsize{8.6bp}{9.03bp}\selectfont]
  at (\FigThreeGuideX,.40356) {$\lambda_*/\pi$\\$\simeq0.2310$};
\node[anchor=north,inner sep=0pt,font=\FigThreeLabelFont]
  at (2.9272,.24688) {Modulus $\lambda/\pi$};

\node[anchor=west,inner sep=0pt,font=\FigThreeLabelFont]
  at (.084375,3.168) {$\mathcal I_\lambda$ spectrum:};
\begin{scope}[xshift=\FigThreeLegendShiftX in,yshift=\FigThreeLegendShiftY in]
\foreach \column/\row/\FigThreeKeyColor/\FigThreeKeyLabel in {
  \FigThreeColumnOne/0/FigThreeBlue/{$0$ (pair)},
  \FigThreeColumnOne/1/FigThreeOrange/{$1/7$ (pair)},
  \FigThreeColumnOne/2/FigThreePurple/{$3/7$ (real)},
  \FigThreeColumnTwo/0/FigThreeGreen/{$0$ (real)},
  \FigThreeColumnTwo/1/FigThreeCyan/{$1/7$ (real)},
  \FigThreeColumnTwo/2/FigThreeGold/{Higher branch}}
{
  \pgfmathsetmacro{\FigThreeRowY}{\FigThreeLegendTop-\row*\FigThreeLegendRowGap}
  \draw[\FigThreeKeyColor,line width=\FigThreeCurveWidth]
    (\column,\FigThreeRowY)--++(\FigThreeLegendLineLength,0);
  \node[anchor=west,inner sep=0pt,font=\FigThreeLegendFont]
    at ({\column+\FigThreeLegendTextOffset},\FigThreeRowY) {\FigThreeKeyLabel};
}
\draw[FigThreeIdentity,line width=\FigThreeCurveWidth,
      dash pattern=on 4.44bp off 1.92bp]
  (\FigThreeColumnThree,\FigThreeLegendTop)--++(\FigThreeLegendLineLength,0);
\node[anchor=west,inner sep=0pt,font=\FigThreeLegendFont]
  at ({\FigThreeColumnThree+\FigThreeLegendTextOffset},\FigThreeLegendTop)
  {$0$ (identity)};
\draw[draw=white,fill=FigThreeEPColor,line width=.4bp]
  plot[only marks,mark=figthreefilledstar,mark size=4bp]
  coordinates {({\FigThreeColumnThree+.5*\FigThreeLegendLineLength},
    {\FigThreeLegendTop-\FigThreeLegendRowGap})};
\node[anchor=west,inner sep=0pt,font=\FigThreeLegendFont]
  at ({\FigThreeColumnThree+\FigThreeLegendTextOffset},
    {\FigThreeLegendTop-\FigThreeLegendRowGap}) {EP};
\end{scope}
\end{tikzpicture}
\endgroup
\caption{Low primary spectrum of $\mathcal I_\lambda$ for $\lambda\in[0,\pi]$. (a) Real parts of the resolved low branches and (b) imaginary parts of the colliding pair. Legend numbers give the weights at $\lambda=0$. 
}
\label{fig:annulus_spectrum}
\end{figure}
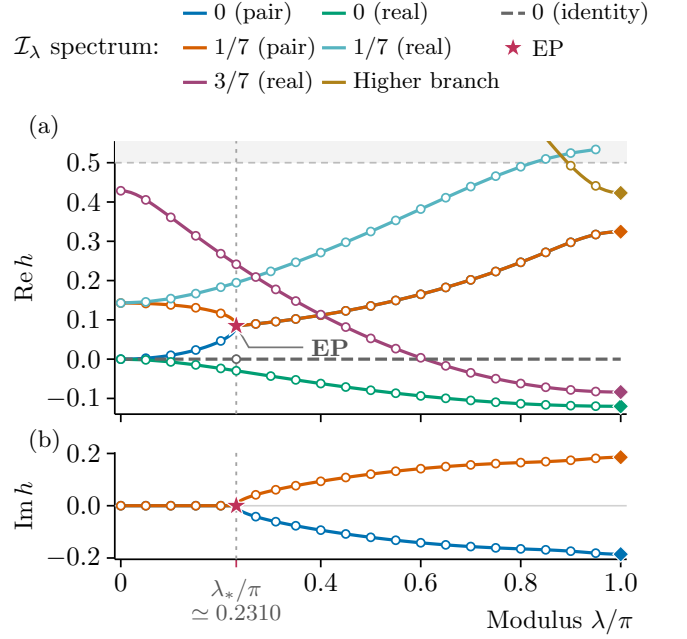

At the factorized point $\lambda=0$, $Z_{\lambda=0}$ can be exactly computed. The vacuum multiplicity $n_{\rm vac}=3$ indicates that $\mathcal I_{\lambda=0}$ is a sum of three simple branes. One can easily show that $\left|\mathcal I_\mathbf v;0\right\rangle=|\one\rangle_L\otimes|\one\rangle_{\overline R}\,+
    |x\rangle_L\otimes|x\rangle_{\overline R}\,+
    |y\rangle_L\otimes|y\rangle_{\overline R}$,
where the three tensor product Cardy branes precisely correspond to our three phantom sectors Eq.~\eqref{eq:phantom_sectors}. We collect all $6$ primary spectra and their multiplicities below $h=\frac12$: a triplet of $h=0$ states, a doublet of $h=\frac17$ states, and a single $h=\frac37$ state, see FIG.~\ref{fig:annulus_spectrum}.

In contrast, turning on the modulus parameter $\lambda$, the interface $\mathcal I_\lambda$ becomes non-factorizable, and the original spectra start to flow along $\lambda$. Among them, the three degenerate states of $h=0$ are resolved: One identity state remains at zero implying the simpleness of $\mathcal I_{\lambda\neq0}$, whereas the other two zero-weight operators move in opposite directions. One acquires a negative weight, and the other approaches an additional operator that flows from one of the degenerate states of $h=\frac{1}{7}$ at $\lambda=0$. The spectral collision between the two operators from the branches of $h=0,\,\frac17$ is quite interesting, because, after collision, they turn out to be a pair of operators with complex conjugate weights $h_{\pm}(\lambda)$. At the maximal transmission point $\lambda=\pi$, the pair of weights are $ h_\pm(\pi)\simeq 0.3244\pm0.1856\,\ii$. As one can show that the self-annulus spectrum of $\mathcal I_\lambda$ is symmetric about $\lambda=\pi$, see more discussions in~\cite{ChenRGCompanion}, we here draw the spectral flow along the modulus parameter $\lambda\in[0,\pi]$ for low weight primaries of $h<\frac12$, in FIG.~\ref{fig:annulus_spectrum} .  

\begin{figure}[t]
\centering
\begingroup
\definecolor{DeltaBlue}{HTML}{0072B2}
\definecolor{DeltaOrange}{HTML}{D55E00}
\definecolor{DeltaInk}{HTML}{303030}
\definecolor{DeltaStar}{HTML}{BF3159}
\def\DeltaLegendY{1.80}       
\def\DeltaAnnulusX{0.98}      
\def\DeltaLinearX{1.78}       
\def\DeltaEPLabelDX{0.0016}   
\def\DeltaEPLabelY{0.62}      
\def\DeltaPointSize{2.05pt}
\def\DeltaCurveWidth{1.15pt}
\def\DeltaLambdaEP{0.7257133012899033}
\def\DeltaEPX{0.23100171833565211}
\def\DeltaEPY{-9.4739363757121993e-09}        
\def\DeltaSlope{-0.013381571399223779}
\pgfdeclareplotmark{DeltaFilledStar}{%
  \pgfpathmoveto{\pgfqpointpolar{90}{\pgfplotmarksize}}%
  \pgfpathlineto{\pgfqpointpolar{126}{0.381966\pgfplotmarksize}}%
  \pgfpathlineto{\pgfqpointpolar{162}{\pgfplotmarksize}}%
  \pgfpathlineto{\pgfqpointpolar{198}{0.381966\pgfplotmarksize}}%
  \pgfpathlineto{\pgfqpointpolar{234}{\pgfplotmarksize}}%
  \pgfpathlineto{\pgfqpointpolar{270}{0.381966\pgfplotmarksize}}%
  \pgfpathlineto{\pgfqpointpolar{306}{\pgfplotmarksize}}%
  \pgfpathlineto{\pgfqpointpolar{342}{0.381966\pgfplotmarksize}}%
  \pgfpathlineto{\pgfqpointpolar{18}{\pgfplotmarksize}}%
  \pgfpathlineto{\pgfqpointpolar{54}{0.381966\pgfplotmarksize}}%
  \pgfpathclose\pgfusepathqfillstroke
}
\begin{tikzpicture}[x=1in,y=1in]
\path[use as bounding box] (0,0) rectangle (3.375,1.94);
\begin{axis}[
  at={(0.597375in,0.405in)},anchor=south west,
  width=2.676375in,height=1.210in,scale only axis,
  axis lines*=left,axis line style={DeltaInk,line width=0.65pt},
  xmin=0.22638622498598718,xmax=0.23561721168531707,ymin=-2.35,ymax=2.35,
  xtick={0.228,0.230,0.232,0.234},ytick={-2,-1,0,1,2},
  xticklabel style={/pgf/number format/fixed,/pgf/number format/precision=3,
    /pgf/number format/fixed zerofill},scaled ticks=false,
  tick align=outside,tick style={DeltaInk,line width=0.65pt},
  major tick length=3pt,
  tick label style={font=\fontsize{8.5}{10}\selectfont,text=DeltaInk},
  xlabel={Modulus $\lambda/\pi$},ylabel={$10^4\,\Delta$},
  xlabel style={at={(axis description cs:0.5,-0.175)},anchor=north,
    font=\fontsize{9.3}{11}\selectfont,text=DeltaInk},
  ylabel style={at={(axis description cs:-0.145,0.5)},anchor=south,
    font=\fontsize{9.3}{11}\selectfont,text=DeltaInk},
  clip mode=individual,axis on top=true,
]
\path[fill=DeltaBlue!3.5!white] (rel axis cs:0,0) rectangle (axis cs:\DeltaEPX,2.35);
\path[fill=DeltaOrange!3.5!white] (axis cs:\DeltaEPX,-2.35) rectangle (rel axis cs:1,1);
\draw[gray!55,line width=0.6pt] (axis cs:0.22638622498598718,0) -- (axis cs:0.23561721168531707,0);
\draw[gray!65,line width=0.75pt,dash pattern=on 2pt off 3pt]
  (axis cs:\DeltaEPX,-2.35) -- (axis cs:\DeltaEPX,2.35);
\addplot[DeltaInk,line width=\DeltaCurveWidth,no marks,
  domain=0.22638622498598718:0.23561721168531707,samples=2]
  {10000*\DeltaSlope*(pi*x-\DeltaLambdaEP)};
\addplot[only marks,mark=*,mark size=\DeltaPointSize,
  mark options={draw=DeltaBlue,fill=white,line width=0.85pt},
  error bars/.cd,y dir=both,y explicit,error bar style={DeltaBlue,line width=0.7pt},
  error mark options={rotate=90,mark size=1.7pt}]
  table[x=x,y=y,y error minus=em,y error plus=ep] {
x y em ep
0.22654537992907905 1.8589458839925117 3.8004957395119249e-06 0.00036203707443597686
0.22765946453072233 1.396933350619096 2.9229055647735959e-06 9.8738674677331488e-06
0.22877354913236558 0.93310048472439133 0.00017628668076740198 9.2365295665012507e-06
0.22988763373400886 0.46745384066267864 6.4346440171680311e-05 8.6311524594829105e-06
};
\addplot[only marks,mark=*,mark size=\DeltaPointSize,
  mark options={draw=DeltaOrange,fill=white,line width=0.85pt},
  error bars/.cd,y dir=both,y explicit,error bar style={DeltaOrange,line width=0.7pt},
  error mark options={rotate=90,mark size=1.7pt}]
  table[x=x,y=y,y error minus=em,y error plus=ep] {
x y em ep
0.23211580293729536 -0.46925447159843242 1.4508857778468037e-05 7.4293323673347038e-06
0.23322988753893864 -0.94030292599918652 7.2417163421206656e-06 6.8146858825163161e-06
0.23434397214058189 -1.4131387248114933 3.6457020188103081e-06 6.1882427052637259e-06
0.2354580567422252 -1.8877551905413934 1.9552983196316834e-06 5.5514770531213051e-06
};
\draw[black,line width=0.65pt]
  (axis cs:\DeltaEPX,\DeltaEPY)
  -- (axis cs:{\DeltaEPX+0.00046},\DeltaEPLabelY)
  -- (axis cs:{\DeltaEPX+\DeltaEPLabelDX-0.00012},\DeltaEPLabelY);
\node[anchor=west,inner sep=0pt,font=\fontsize{8.7}{10.4}\selectfont\bfseries]
  at (axis cs:{\DeltaEPX+\DeltaEPLabelDX},\DeltaEPLabelY) {EP};
\addplot[only marks,mark=DeltaFilledStar,mark size=4.5pt,
  mark options={draw=white,fill=DeltaStar,line width=0.5pt},
  error bars/.cd,y dir=both,y explicit,error bar style={DeltaStar,line width=0.7pt},
  error mark options={rotate=90,mark size=1.7pt}]
  table[x=x,y=y,y error minus=em,y error plus=ep] {
x y em ep
0.23100171833565211 -9.4739363757121993e-09 2.953428945075874e-05 8.0330514608294645e-06
};
\node[anchor=north west,inner sep=0pt,text=DeltaBlue,font=\fontsize{8.5}{10}\selectfont]
  at (rel axis cs:0.28,0.89) {Real pair};
\node[anchor=south,inner sep=0pt,text=DeltaOrange,font=\fontsize{8.5}{10}\selectfont]
  at (rel axis cs:0.67,0.09) {Complex pair};
\end{axis}
\draw[draw=black!65,fill=white,line width=0.85pt]
  (\DeltaAnnulusX-.25,\DeltaLegendY) circle[radius=1.85pt];
\node[anchor=west,inner sep=0pt,text=DeltaInk,font=\fontsize{8}{9.6}\selectfont]
  at ({\DeltaAnnulusX-.25+0.12},\DeltaLegendY) {Self-annulus};
\draw[DeltaInk,line width=1.15pt]
  (\DeltaLinearX,\DeltaLegendY) -- ({\DeltaLinearX+0.15},\DeltaLegendY);
\node[anchor=west,inner sep=0pt,text=DeltaInk,font=\fontsize{8}{9.6}\selectfont]
  at ({\DeltaLinearX+0.20},\DeltaLegendY) {Linear approximation};
\end{tikzpicture}
\endgroup
\caption{Squared gap of the pair of weights starting at $h=0$ and $\frac17$. Circles show equally spaced annulus measurements, with the star marking a direct evaluation at the estimated EP angle. The solid line is a linear fit of $\Delta(\lambda)$ near $\lambda_*$}
\label{fig:annulus-delta}
\end{figure}
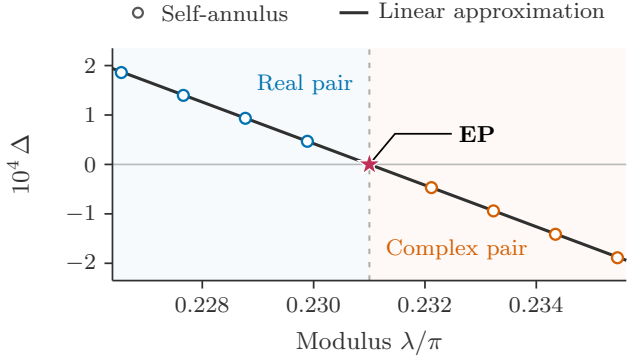

The collision point in the moduli space is located at
\begin{align}
    \lambda_*/\pi\simeq0.23100\,,\qquad h_*\simeq0.08477\,,
    \label{eq:EP_modulus}
\end{align}
To determine the behavior of the pair of weights $h_{\pm}(\lambda)$ near $\lambda_*$, we define their mean and squared gap,
\begin{align}
    m(\lambda)=\frac{h_+(\lambda)+h_-(\lambda)}{2}\,,\quad \Delta(\lambda)=\frac{(h_+(\lambda)-h_-(\lambda))^2}{4}\,.
    \label{eq:m_Delta}
\end{align}
FIG.~\ref{fig:annulus-delta} shows self-annulus measurements on both sides of the collision. Fitting the data, we find a linear simple zero for $\Delta(\lambda)$ near $\lambda_*$, 
\begin{align}
    \Delta(\lambda)=\kappa(\lambda-\lambda_*)+\mathcal O\left((\lambda-\lambda_*)^2\right)\,,\ \ \kappa\simeq-0.01338
    \label{eq:kappa_slope}
\end{align}
It signals that the collision point~\eqref{eq:EP_modulus} in the moduli space is an \emph{exceptional point} (EP), where the boundary Hamiltonian cannot be diagonalized. Instead, the Hamiltonian should be written as a rank-two Jordan block in the basis of the pair of operators, one of which then exhibits logarithmic scaling behavior~\cite{Heiss:2012dx}.

\emph{Logarithmic operators at the collision.---}
Having identified the EP from the self-annulus, we now aim to determine its logarithmic conformal block from a physical correlator. To track the behaviors of the two operators at the collision, we introduce a bulk probe $\Phi$ in $\mathcal T_A$ from Eq.~\eqref{eq:duality}, $\Phi=\left(\one_{\rm Ising}\otimes\phi_{1,3}\right)\otimes\overline{\one}_{R}$, where $\phi_{1,3}$ is the primary in $\mathcal M_{4,7}$ with weight $\frac17$. Notice that the fusion rule $\phi_{1,3}\times\phi_{1,3}=\phi_{1,1}+\phi_{1,3}+\phi_{1,5}$ has output channels of weights $0,\,\frac17$ and $\frac{10}{7}$. Therefore, in the mirror picture, the chiral component of $\phi_{1,3}^{\mathcal M_{4,7}}$ and its anti-chiral mirror can provide fusion channels to the two target boundary operators in the $0$ and $\frac17$ branches at $\lambda=0$. Now inserting the probes $\Phi$ at origin and $r$ of a unit disk with boundary condition $\mathcal I_\lambda$, we define the probe's two-point  correlator in the presence of $\mathcal I_\lambda$,
\begin{align}
G_\lambda(r)
=\frac{\left\langle \mathcal I_\mathbf v;\lambda\right|\Phi(r,r)\Phi(0,0)\left|0\right\rangle}
{\left\langle \mathcal I_\mathbf v;\lambda\big|0\right\rangle}\,,
\qquad 0<r<1\,.
\end{align}
In the mirror picture, denoting $V$ and $\widetilde V$ the chiral component of probe $\Phi$ and its anti-chiral mirror, we rewrite the two-point correlator $G_\lambda(r)$ as
\begin{align}
G_\lambda(\sqrt{z})
=
\left\langle
V(0)V(z)\widetilde V(1)\widetilde V(\infty)
\right\rangle^{\mathrm{mirror}}_{\mathcal I_\lambda},
\ z=r^2\,,
\end{align}
where the subscript $\mathcal I_\lambda$ specifies that the chiral and anti-chiral operators are connected with the boundary $\mathcal I_\lambda$. As the modulus $\lambda$ varies, the conformal block data, especially for the logarithmic block at the collision, of the two target boundary operators in the $t$-channel can be reconstructed from the bulk correlator in the $s$-channel via the bulk-boundary crossing equations~\cite{Cardy:1991tv,Lewellen:1991tb}, 
\begin{align}
G_\lambda(\sqrt z)
 =\sum_{\ell}\mu_{\ell}(\lambda)
 \cF_{h_{\ell}(\lambda)}(1-z)\,,
\label{eq:crossing}
\end{align}
where $\ell$ labels the boundary conformal families coupled to the probe, $\cF_h$ is the boundary Virasoro block of weight $h$, and $\mu_\ell$ is the corresponding bulk-boundary OPE coefficient. These OPE coefficients and conformal weights need not be positive and may be complex.

\begin{figure}[t]
\centering
\begingroup
\def\FigFiveCanvasWidth{3.375}
\def\FigFiveCanvasHeight{3.40}
\def\FigFiveAxesLeft{0.57375}
\def\FigFiveAxesWidth{2.70}
\def\FigFiveRealBottom{1.496}
\def\FigFiveRealHeight{1.309}
\def\FigFiveImagBottom{0.578}
\def\FigFiveImagHeight{0.697}
\def\FigFiveInsetLeft{0.95175}
\def\FigFiveInsetBottom{2.340305}
\def\FigFiveInsetWidth{1.161}
\def\FigFiveInsetHeight{0.425425}
\def\FigFiveLegendShiftX{0}
\def\FigFiveLegendShiftY{0}
\def\FigFiveEP{0.23100146327715418}
\def\FigFiveCurveWidth{1.40bp}
\def\FigFiveInsetCurveWidth{1.05bp}
\def\FigFiveCircleRadius{1.50bp}
\def\FigFiveInsetCircleRadius{1.25bp}
\def\FigFiveStarRadius{3.45bp}
\def\FigFiveInsetStarRadius{2.90bp}
\def\FigFiveLabelFont{\fontsize{9.5bp}{11.4bp}\selectfont}
\def\FigFiveTickFont{\fontsize{9bp}{10.8bp}\selectfont}
\def\FigFiveLegendFont{\fontsize{8.7bp}{10.4bp}\selectfont}
\def\FigFiveInsetTickFont{\fontsize{7.6bp}{9.1bp}\selectfont}
\definecolor{FigFiveBlue}{HTML}{0072B2}
\definecolor{FigFiveOrange}{HTML}{D55E00}
\definecolor{FigFiveInk}{HTML}{303030}
\definecolor{FigFiveGuide}{HTML}{929292}
\definecolor{FigFiveCircle}{HTML}{555555}
\definecolor{FigFiveInsetFrame}{HTML}{777777}
\definecolor{FigFiveEPColor}{HTML}{BF3159}

\pgfdeclareplotmark{figfivefilledstar}{%
  \pgfpathmoveto{\pgfpointpolar{90}{\pgfplotmarksize}}%
  \pgfpathlineto{\pgfpointpolar{126}{0.38196601125\pgfplotmarksize}}%
  \pgfpathlineto{\pgfpointpolar{162}{\pgfplotmarksize}}%
  \pgfpathlineto{\pgfpointpolar{198}{0.38196601125\pgfplotmarksize}}%
  \pgfpathlineto{\pgfpointpolar{234}{\pgfplotmarksize}}%
  \pgfpathlineto{\pgfpointpolar{270}{0.38196601125\pgfplotmarksize}}%
  \pgfpathlineto{\pgfpointpolar{306}{\pgfplotmarksize}}%
  \pgfpathlineto{\pgfpointpolar{342}{0.38196601125\pgfplotmarksize}}%
  \pgfpathlineto{\pgfpointpolar{18}{\pgfplotmarksize}}%
  \pgfpathlineto{\pgfpointpolar{54}{0.38196601125\pgfplotmarksize}}%
  \pgfpathclose\pgfusepathqfillstroke
}

\pgfplotstableread[row sep=\\]{
u rminus rplus iminus iplus \\
0 -7.4672741490177282e-44 0.14285714285714285 0 0 \\
0.031830988618379068 0.0009660071299580664 0.14239550862875258 0 0 \\
0.063661977236758135 0.0039010887133269453 0.14097400290365206 0 0 \\
0.095492965855137196 0.0089289216269616469 0.13847035548795467 0 0 \\
0.12732395447351627 0.016308274565079058 0.13462825755290969 0 0 \\
0.15915494309189535 0.026572131966908972 0.12891841041695859 0 0 \\
0.19098593171027439 0.041021136527914143 0.12004528639159569 0 0 \\
0.20690142601946396 0.050975908908253202 0.11326371949125162 0 0 \\
0.22281692032865347 0.065418728395162246 0.10225228714845394 0 0 \\
0.22345354010102106 0.066210772000650606 0.10160288528549763 0 0 \\
0.22409015987338862 0.067035427949533918 0.100921286103164 0 0 \\
0.22472677964575621 0.067897224574799298 0.10020296134844803 0 0 \\
0.22500000000000001 0.06827984080974174 0.09988204676723271 0 0 \\
0.22536339941812381 0.068801870864450429 0.099442202113707731 0 0 \\
0.22600001919049137 0.069756739163321757 0.098631636134856565 0 0 \\
0.22663663896285896 0.070771635973407201 0.097761456991050516 0 0 \\
0.22727325873522652 0.07186010392695602 0.096818122131642259 0 0 \\
0.22790987850759412 0.073041785502305348 0.095781989160580144 0 0 \\
0.22854649827996171 0.074347149071748636 0.094622589791034439 0 0 \\
0.22918311805232927 0.075828325177449585 0.093287793533458035 0 0 \\
0.22981973782469686 0.077589866138160921 0.091673048208531269 0 0 \\
0.23000000000000001 0.078168877504785711 0.091135678322538438 0 0 \\
0.23045635759706445 0.0799195927510402 0.089490533064400093 0 0 \\
0.23052001957430121 0.080214751618024271 0.089210118218685824 0 0 \\
0.23058368155153799 0.080529702013132753 0.08890991600416924 0 0 \\
0.23064734352877472 0.080869174106211472 0.088585196251089265 0 0 \\
0.23071100550601148 0.08124015166148367 0.088228975195307313 0 0 \\
0.23077466748324824 0.08165374213852887 0.087830145377328472 0 0 \\
0.23083832946048499 0.082129693274264282 0.087368959060320175 0 0 \\
0.23090199143772178 0.08271040288973476 0.086803018423322123 0 0 \\
0.23096565341495853 0.083534590693813762 0.085993603757545431 0 0 \\
0.23098000000000002 0.083812544671877293 0.085718979563741679 0 0 \\
0.23098156890926771 0.083847972441063165 0.085683915944854916 0 0 \\
0.2309974844035769 0.08435111250134314 0.085184470079129812 0 0 \\
0.23100000000000001 0.084507568496659885 0.085028598010465972 0 0 \\
0.23100066750243872 0.084568644764765097 0.08496767668581845 0 0 \\
0.23100130412221112 0.084654026368058397 0.084882442857795268 0 0 \\
0.23100146327715418 0.084688549006039615 0.084847957163696572 0 0 \\
0.23100162243209729 0.084768271556822358 0.084768271556822358 -1.8386041086888476e-05 1.8386041086888476e-05 \\
0.23100194074198346 0.084768308500769884 0.084768308500769884 -0.00011713082817291433 0.00011713082817291433 \\
0.23100385060130058 0.084768530165547029 0.084768530165547029 -0.00030660959866438023 0.00030660959866438023 \\
0.23101000000000002 0.084769243898421748 0.084769243898421748 -0.00059374371819299995 0.00059374371819299995 \\
0.23101339989788608 0.084769638517512555 0.084769638517512555 -0.00070389378365544157 0.00070389378365544157 \\
0.23102931539219529 0.084771485874787944 0.084771485874787944 -0.0010791621176367773 0.0010791621176367773 \\
0.23105000000000003 0.084773886991074784 0.084773886991074784 -0.0014262800677797244 0.0014262800677797244 \\
0.23109297736943202 0.084778876603895797 0.084778876603895797 -0.0019599632558857842 0.0019599632558857842 \\
0.23115663934666877 0.084786269413045429 0.084786269413045429 -0.0025532153903748875 0.0025532153903748875 \\
0.23122030132390556 0.084793664302279123 0.084793664302279123 -0.0030326472509175041 0.0030326472509175041 \\
0.23128396330114231 0.08480106127163918 0.08480106127163918 -0.0034460959646571945 0.0034460959646571945 \\
0.23134762527837907 0.084808460321167914 0.084808460321167914 -0.0038150744230826673 0.0038150744230826673 \\
0.23141128725561583 0.084815861450907623 0.084815861450907623 -0.0041514570446040452 0.0041514570446040452 \\
0.23172959714179961 0.084852898304252911 0.084852898304252911 -0.0055355590704240813 0.0055355590704240813 \\
0.2323662169141672 0.084927128051138515 0.084927128051138515 -0.0075830090794688611 0.0075830090794688611 \\
0.23300283668653476 0.08500156588704405 0.08500156588704405 -0.0091880449300868576 0.0091880449300868576 \\
0.23363945645890236 0.085076211854639164 0.085076211854639164 -0.010554464408863639 0.010554464408863639 \\
0.23427607623126995 0.0851510659967816 0.0851510659967816 -0.011765693785734766 0.011765693785734766 \\
0.23491269600363751 0.085226128356521186 0.085226128356521186 -0.012865634023085592 0.012865634023085592 \\
0.23500000000000001 0.085236438405565232 0.085236438405565232 -0.013009400914617192 0.013009400914617192 \\
0.2355493157760051 0.085301398977101831 0.085301398977101831 -0.013880762211608583 0.013880762211608583 \\
0.23618593554837269 0.085376877901962647 0.085376877901962647 -0.014828502525894977 0.014828502525894977 \\
0.23682255532074026 0.085452565174738898 0.085452565174738898 -0.01572104309101606 0.01572104309101606 \\
0.23745917509310785 0.085528460839262688 0.085528460839262688 -0.016567303935019661 0.016567303935019661 \\
0.23809579486547544 0.085604564939563602 0.085604564939563602 -0.017374045361559825 0.017374045361559825 \\
0.238732414637843 0.085680877519869303 0.085680877519869303 -0.018146535395038899 0.018146535395038899 \\
0.25464790894703254 0.08765658151510021 0.08765658151510021 -0.032148991171374333 0.032148991171374333 \\
0.28647889756541162 0.092002018863440282 0.092002018863440282 -0.050409481266119133 0.050409481266119133 \\
0.31830988618379069 0.096878733206169548 0.096878733206169548 -0.064541685950080363 0.064541685950080363 \\
0.35014087480216977 0.10229512662032421 0.10229512662032421 -0.076729109664218864 0.076729109664218864 \\
0.38197186342054879 0.10826124521523117 0.10826124521523117 -0.087663148001018737 0.087663148001018737 \\
0.41380285203892792 0.11478892419528494 0.11478892419528494 -0.097647518739712502 0.097647518739712502 \\
0.44563384065730693 0.12189191752898464 0.12189191752898464 -0.10683488811550174 0.10683488811550174 \\
0.47746482927568601 0.12958600334678663 0.12958600334678663 -0.11530622109958728 0.11530622109958728 \\
0.50929581789406508 0.13788904751016606 0.13788904751016606 -0.12310439389508013 0.12310439389508013 \\
0.54112680651244416 0.14682099137674276 0.14682099137674276 -0.13025097403429456 0.13025097403429456 \\
0.57295779513082323 0.15640370226620642 0.15640370226620642 -0.1367560132214084 0.1367560132214084 \\
0.60478878374920231 0.16666058320914925 0.16666058320914925 -0.14262499906054615 0.14262499906054615 \\
0.63661977236758138 0.17761577993798988 0.17761577993798988 -0.14786517570596897 0.14786517570596897 \\
0.66845076098596046 0.18929274596259293 0.18929274596259293 -0.15249279760350845 0.15249279760350845 \\
0.70028174960433953 0.20171181990618536 0.20171181990618536 -0.15654267669881747 0.15654267669881747 \\
0.7321127382227185 0.21488626775692751 0.21488626775692751 -0.16008106554719059 0.16008106554719059 \\
0.76394372684109757 0.22881567546608836 0.22881567546608836 -0.16322163518743038 0.16322163518743038 \\
0.79577471545947676 0.24347375611385005 0.24347375611385005 -0.16614041641863334 0.16614041641863334 \\
0.82760570407785583 0.25878220620466819 0.25878220620466819 -0.16907652257935304 0.16907652257935304 \\
0.85943669269623491 0.27454869503368512 0.27454869503368512 -0.17228999198030553 0.17228999198030553 \\
0.89126768131461387 0.29032332824957108 0.29032332824957108 -0.1759358035333235 0.1759358035333235 \\
0.92309866993299294 0.30512285982993737 0.30512285982993737 -0.17984428073382316 0.17984428073382316 \\
0.95492965855137202 0.31711225789671954 0.31711225789671954 -0.18333566510213858 0.18333566510213858 \\
0.9867606471697512 0.32376085986576203 0.32376085986576203 -0.18537127861712507 0.18537127861712507 \\
1 0.32442654878808197 0.32442654878808197 -0.18557801559632384 0.18557801559632384 \\
}\FigFiveCrossData

\pgfplotstableread[row sep=\\]{
u rminus rplus iminus iplus \\
0 0 0.14285714285714285 0 0 \\
0.050000000000000003 0.0023945314402931857 0.14170723616110489 -4.8819999036950551e-11 -2.7235390832441262e-09 \\
0.10000000000000001 0.0098225116316045125 0.13801598745145918 -3.4545046122540678e-12 -1.6206277418483288e-10 \\
0.14999999999999999 0.023273315265705603 0.13080302824414611 -1.3115937488052255e-09 -4.5213788494898755e-08 \\
0.20000000000000001 0.046326322249874825 0.11650572034584313 -7.7722868766818105e-12 -2.0800776514323877e-10 \\
0.27000000000000002 0.089686565959899822 0.08968656595244906 -0.041771692710493975 0.041771692716663235 \\
0.29999999999999999 0.094008227263091035 0.094008227244328724 -0.056727275133822253 0.056727275137044023 \\
0.34999999999999998 0.10227006063451582 0.10227006064824185 -0.076678148242347394 0.076678148254950812 \\
0.40000000000000002 0.11188882192772533 0.11188882192924936 -0.093422285842796285 0.09342228583326817 \\
0.45000000000000001 0.12291275744611278 0.12291275771231408 -0.10803678275862733 0.10803678259055746 \\
0.5 0.13540268541863185 0.13540268563294436 -0.12089119834426712 0.12089119833495265 \\
0.55000000000000004 0.14943366269016642 0.14943366285808729 -0.13211666802105113 0.13211666809300435 \\
0.59999999999999998 0.16509710059286964 0.16509710016033091 -0.14174761874167754 0.14174761809893918 \\
0.65000000000000002 0.18250336050943899 0.18250336096402883 -0.14978580520072526 0.14978580867629832 \\
0.69999999999999996 0.20178262322356599 0.20178262355566984 -0.15624987663458123 0.15624987829900427 \\
0.75 0.22307849555895348 0.22307849539714883 -0.16124572407190571 0.16124572698369036 \\
0.80000000000000004 0.24649435838174 0.24649435812413018 -0.16511250863714771 0.16511250954249379 \\
0.84999999999999998 0.27182566457496765 0.271825662282492 -0.1687409701435181 0.16874097196245488 \\
0.90000000000000002 0.29737629824557804 0.29737629347533584 -0.17394446339069025 0.17394446492707066 \\
0.94999999999999996 0.31726968973106717 0.31726971523761222 -0.18144378781816756 0.18144379241943254 \\
1 0.32435589180949642 0.32435554337743194 -0.18562756230102098 0.1856273757286136 \\
}\FigFiveAnnData

\pgfplotstableread[row sep=\\]{
u rminus rplus iminus iplus \\
0.22 0.06222403314407788 0.10482079104999423 -2.0575804484465959e-10 1.5258277471445849e-09 \\
0.2225 0.065035339025302605 0.10256487118624392 -8.8395887707725221e-11 1.830549628486975e-09 \\
0.22500000000000001 0.068279728207177287 0.099882224993815424 -1.586283841466115e-10 1.8539008037252001e-09 \\
0.23000000000000001 0.078168546166649311 0.0911360983909629 -1.3842612638365981e-09 3.3885875468230419e-09 \\
0.23499999999999999 0.085236500415922165 0.085236502718994661 -0.013009201142602103 0.01300920350825813 \\
0.23999999999999999 0.085833575642213569 0.085833575647975099 -0.019598360862803411 0.019598360549100864 \\
}\FigFiveInsetData

\pgfplotstableread[row sep=\\]{
u rminus rplus iminus iplus \\
0.23100146327715418 0.084688549006039615 0.084847957163696572 0 0 \\
}\FigFiveEPData


\begin{tikzpicture}[x=1in,y=1in,text=FigFiveInk]
\path[use as bounding box] (0,0) rectangle
  (\FigFiveCanvasWidth,\FigFiveCanvasHeight);
\pgfplotsset{
  fig five main/.style={
    scale only axis,
    width=\FigFiveAxesWidth in,
    xmin=-0.012,xmax=1.012,
    axis x line*=bottom,axis y line*=left,
    axis line style={draw=FigFiveInk,line width=.65bp},
    tick align=outside,major tick length=3bp,
    tick style={draw=FigFiveInk,line width=.65bp},
    tick label style={font=\FigFiveTickFont,text=FigFiveInk},
    scaled ticks=false,
    xtick={0,.4,.6,.8,1},
    xticklabels={$0$,$0.4$,$0.6$,$0.8$,$1.0$},
    xticklabel style={yshift=1bp},
    yticklabel style={xshift=1bp},
    axis background/.style={fill=white},
    clip=true,
  },
  fig five annulus/.style={
    only marks,mark=*,mark size=\FigFiveCircleRadius,
    mark options={solid,draw=FigFiveCircle,fill=white,line width=.60bp},
  },
  fig five stars/.style={
    only marks,mark=figfivefilledstar,mark size=\FigFiveStarRadius,
    mark options={solid,draw=white,fill=FigFiveEPColor,line width=.35bp},
  },
}

\begin{axis}[
  fig five main,
  at={(\FigFiveAxesLeft in,\FigFiveRealBottom in)},anchor=south west,
  height=\FigFiveRealHeight in,
  ymin=-.014,ymax=.356,
  ytick={0,.1,.2,.3},yticklabels={$0.0$,$0.1$,$0.2$,$0.3$},
  xticklabels=\empty,
]
\draw[FigFiveGuide,line width=.7bp,dash pattern=on 2bp off 3bp]
  (axis cs:\FigFiveEP,-.014)--(axis cs:\FigFiveEP,.356);
\addplot[FigFiveBlue,line width=\FigFiveCurveWidth,no marks]
  table[x=u,y=rminus]{\FigFiveCrossData};
\addplot[FigFiveOrange,line width=\FigFiveCurveWidth,no marks]
  table[x=u,y=rplus]{\FigFiveCrossData};
\addplot[fig five annulus] table[x=u,y=rminus]{\FigFiveAnnData};
\addplot[fig five annulus] table[x=u,y=rplus]{\FigFiveAnnData};
\addplot[fig five stars] table[x=u,y=rminus]{\FigFiveEPData};
\addplot[fig five stars] table[x=u,y=rplus]{\FigFiveEPData};
\draw[FigFiveInsetFrame,line width=.7bp]
  (axis cs:\FigFiveEP,.076)--(axis cs:.2587,.041)--(axis cs:.35,.041);
\node[anchor=west,inner sep=0pt,text=FigFiveCircle,
      font=\fontsize{9bp}{10.8bp}\selectfont\bfseries]
  at (axis cs:.37,.041) {EP};
\end{axis}

\begin{axis}[
  fig five main,
  at={(\FigFiveAxesLeft in,\FigFiveImagBottom in)},anchor=south west,
  height=\FigFiveImagHeight in,
  ymin=-.215,ymax=.215,
  ytick={-.2,0,.2},yticklabels={$-0.2$,$0.0$,$0.2$},
  minor xtick={\FigFiveEP},minor tick length=3bp,
]
\draw[black!18,line width=.6bp] (axis cs:-.012,0)--(axis cs:1.012,0);
\draw[FigFiveGuide,line width=.7bp,dash pattern=on 2bp off 3bp]
  (axis cs:\FigFiveEP,-.215)--(axis cs:\FigFiveEP,.215);
\addplot[FigFiveBlue,line width=\FigFiveCurveWidth,no marks]
  table[x=u,y=iminus]{\FigFiveCrossData};
\addplot[FigFiveOrange,line width=\FigFiveCurveWidth,no marks]
  table[x=u,y=iplus]{\FigFiveCrossData};
\addplot[fig five annulus] table[x=u,y=iminus]{\FigFiveAnnData};
\addplot[fig five annulus] table[x=u,y=iplus]{\FigFiveAnnData};
\addplot[fig five stars] table[x=u,y=iminus]{\FigFiveEPData};
\addplot[fig five stars] table[x=u,y=iplus]{\FigFiveEPData};
\end{axis}

\begin{axis}[
  scale only axis,
  at={(\FigFiveInsetLeft in,\FigFiveInsetBottom in)},anchor=south west,
  width=\FigFiveInsetWidth in,height=\FigFiveInsetHeight in,
  xmin=.2196,xmax=.2404,ymin=.0605,ymax=.1095,
  axis lines=box,
  xtick pos=bottom,ytick pos=left,
  axis line style={draw=FigFiveInsetFrame,line width=.55bp},
  axis background/.style={fill=white},
  tick align=outside,major tick length=2bp,
  tick style={draw=FigFiveInk,line width=.5bp},
  tick label style={font=\FigFiveInsetTickFont,text=FigFiveInk},
  xtick={.22,.23,.24},xticklabels={$0.22$,$0.23$,$0.24$},
  ytick={.07,.10},yticklabels={$0.07$,$0.10$},
  xticklabel style={yshift=2bp},yticklabel style={xshift=2bp},
  scaled ticks=false,clip=true,
]
\draw[FigFiveGuide,line width=.65bp,dash pattern=on 2bp off 3bp]
  (axis cs:\FigFiveEP,.0605)--(axis cs:\FigFiveEP,.1095);
\addplot[FigFiveBlue,line width=\FigFiveInsetCurveWidth,no marks]
  table[x=u,y=rminus]{\FigFiveCrossData};
\addplot[FigFiveOrange,line width=\FigFiveInsetCurveWidth,no marks]
  table[x=u,y=rplus]{\FigFiveCrossData};
\addplot[fig five annulus,mark size=\FigFiveInsetCircleRadius]
  table[x=u,y=rminus]{\FigFiveInsetData};
\addplot[fig five annulus,mark size=\FigFiveInsetCircleRadius]
  table[x=u,y=rplus]{\FigFiveInsetData};
\addplot[fig five stars,mark size=\FigFiveInsetStarRadius]
  table[x=u,y=rminus]{\FigFiveEPData};
\addplot[fig five stars,mark size=\FigFiveInsetStarRadius]
  table[x=u,y=rplus]{\FigFiveEPData};
\node[anchor=north east,inner sep=0pt,text=FigFiveCircle,
      font=\fontsize{7.5bp}{9bp}\selectfont]
  at (rel axis cs:.95,.91) {EP};
\end{axis}

\node[rotate=90,inner sep=0pt,font=\FigFiveLabelFont]
  at (.1295,2.1505) {$\mathrm{Re}\,h$};
\node[rotate=90,inner sep=0pt,font=\FigFiveLabelFont]
  at (.1295,.9265) {$\mathrm{Im}\,h$};
\node[anchor=base west,inner sep=0pt,font=\FigFiveTickFont]
  at (.11475,2.84427) {(a)};
\node[anchor=base west,inner sep=0pt,font=\FigFiveTickFont]
  at (.11475,1.30985) {(b)};
\pgfmathsetmacro{\FigFiveGuideX}
  {\FigFiveAxesLeft+(\FigFiveEP+.012)/1.024*\FigFiveAxesWidth}
\node[anchor=north,inner sep=0pt,align=center,text=FigFiveCircle,
      font=\fontsize{8.3bp}{8.3bp}\selectfont]
  at (\FigFiveGuideX,.45254) {$\lambda_*/\pi$\\$\simeq0.2310$};
\node[anchor=north,inner sep=0pt,font=\FigFiveLabelFont]
  at (2.91775,.3374) {Modulus $\lambda/\pi$};

\node[anchor=west,inner sep=0pt,font=\FigFiveLabelFont]
  at (.16875,3.179) {$h=0$ and $\frac{1}{7}$ branches:};
\begin{scope}[xshift=\FigFiveLegendShiftX in,yshift=\FigFiveLegendShiftY in]
\draw[FigFiveBlue,line width=1.4bp]
  (1.72125,3.2907708)--(1.8058333,3.2907708);
\draw[FigFiveOrange,line width=1.4bp]
  (1.848125,3.2907708)--(1.9327083,3.2907708);
\draw[draw=white,fill=FigFiveEPColor,line width=.25bp]
  plot[only marks,mark=figfivefilledstar,mark size=3bp]
  coordinates {(2.0172917,3.2907708)};
\node[anchor=base west,inner sep=0pt,font=\FigFiveLegendFont]
  at (2.1260417,3.2484792) {Two-point correlator};
\draw[draw=FigFiveCircle,fill=white,line width=.6bp]
  (1.8904167,3.0913958) circle[radius=1.5bp];
\node[anchor=base west,inner sep=0pt,font=\FigFiveLegendFont]
  at (2.1260417,3.0491042) {Self-annulus};
\end{scope}

\end{tikzpicture}
\endgroup
\caption{Spectral collision between the $h=0$ and $\frac17$ branches.
Curves connect weights extracted from bulk two-point correlators, while hollow circles are self-annulus results. The inset shows the real branches near the collision, marked by the vertical dotted line.}
\label{fig:correlator_spectrum}
\end{figure}
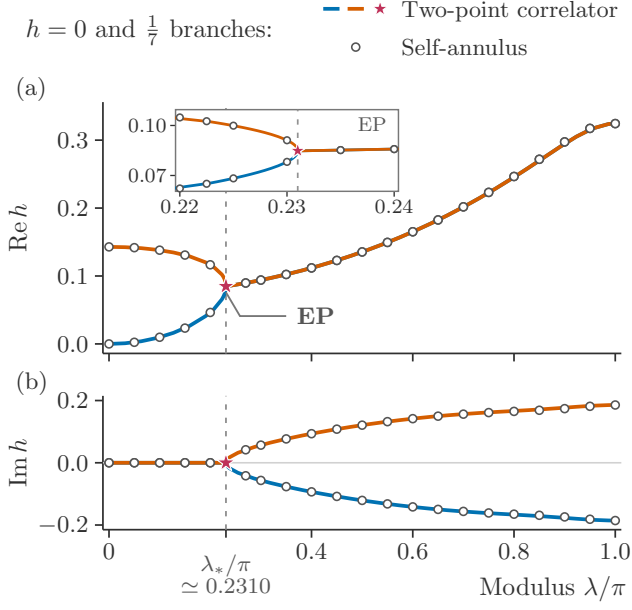

In the $s$-channel, only the vacuum sector of $\mathcal T_A$
contributes to the correlator, as the other sectors in the
OPE of $\Phi(r,r)\Phi(0)$ have vanishing overlap with the boundary state
$\mathcal I_\lambda$. Therefore $G_\lambda(\sqrt z)$ can be easily computed order by order in $z$,
\begin{align}
    &G_\lambda(\sqrt z)=z^{-2h_\Phi}\sum_{N\geq0}a_N(\lambda)z^N\,,\notag\\[2pt]
    &a_N(\lambda)=\left\langle\alpha_N\right|e^{i\lambda Q_0}\left|\alpha_N\right\rangle\,,\quad h_\Phi=\frac17\,,
    \label{eq:s-channel_G}
\end{align}
    where $|\alpha_N\rangle$ are the level-$N$ OPE states embedded in the vacuum of auxiliary theory $\mathcal T_B$ by the topological interface $\mathcal I_{A\to B}$. On the other hand, in the $t$-channel, the OPE coefficients and conformal weights may be complex. The Virasoro blocks have central charge $c_A=-\frac15$, with $\cF_h(\eta)=\eta^{h-2h_\Phi}\left(1+\frac h2\eta+\mathcal O(\eta^2)\right)$ for $\eta=1-z$. We solve Eq.~\eqref{eq:crossing} with a full Virasoro block ansatz in an intermediate $z$-region~\cite{Perlmutter:2015iya}, independently of the results from self-annulus partition functions~\eqref{eq:physical-annulus}. The spectral data agree with those from self-annulus up to $10^{-5}$. The trajectories of operators in the $h=0$ and $\frac17$ branches are reconstructed completely from the two-point correlators $G_\lambda(r)$, as shown in FIG.~\ref{fig:correlator_spectrum}.

Just right at the collision, the two ordinary target blocks $\cF_{h_\pm}$ become identical. We rewrite the regular block as $\mathcal E_{m,\Delta}=\frac{\mathcal F_{m+\delta}+\mathcal F_{m-\delta}}{2}$ and $\mathcal D_{m,\Delta}=\frac{\mathcal F_{m+\delta}-\mathcal F_{m-\delta}}{2\delta}$, where we also relabel the weights $h_\pm$ in terms of $m$ and $\delta\equiv\sqrt \Delta$ defined in Eq.~\eqref{eq:m_Delta}. If the collision is an EP, at $\Delta=0$, one would expect the blocks to become~\cite{Hogervorst:2016itc},
\begin{align}
    G_{\lambda_*}(\sqrt{1-\eta})\sim A_\Phi\mathcal F_{h_*}(\eta)+K_\Phi\frac{\partial \mathcal F_h}{\partial h}\bigg|_{h_*}+\cdots\,.
    \label{eq:crossing_ansatz}
\end{align}
Indeed, at the collision point Eq.~\eqref{eq:EP_modulus}, the crossing equation fits very well for the coefficients $A_\Phi=1.01163$ and $K_\Phi=0.23153$ of the block $\mathcal F_{h_*=0.08477}$. The derivative term producing $\eta^{h_*-2h_{\Phi}}\log\eta$ remains stable under changes in the expansion order in Eq.~\eqref{eq:s-channel_G}, the number of terms retained for higher boundary contributions in Eq.~\eqref{eq:crossing_ansatz} and the fitting $z$-region.

\emph{Structure of the boundary Hamiltonian.---}
The annulus and correlator computations identify the EP in two complementary ways. Let $H(\lambda)$ be the boundary Hamiltonian, restricted to the two dimensional subspace of the colliding operators, with eigenvalues $h_\pm(\lambda)$. In terms of Eq.~\eqref{eq:m_Delta}, we rewrite $H(\lambda)$ as
\begin{align}
    H(\lambda)=m(\lambda)\one_2+K(\lambda),\,\quad{\rm with}\ \ 
    K(\lambda)^2=\Delta(\lambda)\one_2
\end{align}
If the Hamiltonian is diagonalizable at $\lambda_*$, i.e. $H(\lambda_*)=h_*\one_2$, its analyticity will then require $K(\lambda)\simeq \mathcal O(\lambda-\lambda_*)$ and hence $\Delta=\mathcal O\left((\lambda-\lambda_*)^2\right)$. It obviously contradicts Eq.~\eqref{eq:kappa_slope}. Instead, our self-annulus computation strongly supports the conclusion that $K(\lambda_*)$ is nonzero but $K(\lambda_*)^2=\Delta(\lambda_*)\one_2=0$.
It therefore requires a rank-two Jordan block basis for the Hamiltonian~\cite{Heiss:2012dx}. 

In this basis, the two colliding states will satisfy
\begin{align}
H(\lambda_*)|O_1\rangle=h_*|O_1\rangle\,,\quad
H(\lambda_*)|O_2\rangle=h_*|O_2\rangle+|O_1\rangle\,,
\label{eq:jordan}
\end{align}
where $O_1$ is the ordinary operator and $O_2$ is its logarithmic partner. Our correlator probes the logarithmic behavior of $O_2$ directly. Two ordinary degenerate primaries would contribute only to the coefficient of $\cF_{h_*}$, not to its derivative as in Eq.~\eqref{eq:crossing_ansatz}. In contrast, a radial propagation of Hamiltonian $H(\lambda_*)=h_*\one_2+K(\lambda_*)$ gives
\begin{align}
\eta^{H_*}
 =\eta^{h_*}
 \left(\one_2+K(\lambda_*)\log\eta\right)\,,
\label{eq:jordan-propagation}
\end{align}
where the $K(\lambda_*)=K_\Phi\one_2$ is measured from our probe $\Phi$ in Eq.~\eqref{eq:crossing_ansatz}.

\emph{Discussion.---}
Our construction can be applied to broad families of non-unitary RCFTs having spin-one phantom currents. For example, the model we have discussed can be systematically extended to a pair of non-unitary minimal models, $\mathcal M_{q-I,q}$ and $\mathcal M_{q,q+I}$, studied in our companion paper~\cite{ChenRGCompanion}, where $q,\,I$ are coprime integers with $1<I\leq q-2$. The conformal interfaces constructed between them have no continuous moduli as there are only spin-two phantom currents. However, their fermionic extension, from a spin-$\frac12$ phantom current, is given by $\mathcal {SM}_{q-I,q+I}\otimes \psi_1$ containing a free Majorana fermion $\psi_1$. Therefore, tensoring an additional fermion $\psi_2$ will introduce a $\rm U(1)$-current $Q=\ii\psi_1\psi_2$. This $\rm U(1)$ rotation can mix their stress tensors with the weight-two operator $\ii\partial\psi_1\psi_2-\ii\psi_1\partial\psi_2$. Bosonizing back to the original theories, it is equivalent to using Ising to tensor either the left or right factor theory of the pair $\mathcal M_{q-I,q}$ and $\mathcal M_{q,q+I}$ with proper orbifoldings, and the spin-one phantom current is precisely the operator $Q$. For example, one such class of theories are $\mathcal M_{2,2l+1}$ and the diagonal $\mathbb Z_2$ orbifold of $\text{Ising}\otimes\mathcal M_{2l+1,4l}$ for $l\geq 2$. Its smallest member is the Lee-Yang $\mathcal M_{2,5}$ connecting to the diagonal $\mathbb Z_2$ extension of  $\text{Ising}\otimes\mathcal M_{5,8}$ with a shared Fibonacci-like fusion category. These interfaces also need not connect distinct theories. The same extended $\text{Ising}\otimes\mathcal M_{5,8}$ theory can admit continuous non-factorizable interfaces with itself. Moreover, in the End Matter, we construct conformal interfaces with defect conformal manifolds between the non-unitary $\mathcal M_{5,12}$ and the unitary Ising CFT, by condensing higher phantom sectors. All these constructions suggest that exceptional points and the associated logarithmic operators are ubiquitous features of defect conformal manifolds in non-unitary CFTs.

Another interesting generalization is to explore higher dimensional defect conformal manifolds. Their associated conformal interfaces can be constructed when there are several spin-one phantom currents shared by two bulk theories. For example, for theories $\frac{\widehat{\mathfrak {osp}}(1|2)_3}{\widehat{\mathfrak u}(1)_3}$ and $\frac{\widehat{\mathfrak {osp}}(1|2)_1^{\otimes 3}}{\widehat{\mathfrak {osp}}(1|2)_3}$, there are two independent spin-one phantom currents whose zero modes can generate two commuting continuous deformations for the interface in our construction. It would be interesting to understand whether tuning both moduli can lead three or more boundary operators to coalesce and produce Jordan blocks of higher rank, and how the operators transform around these collision loci.

At last, it is also worth mentioning that there are recent developments on related topics, such as complex defect conformal manifolds~\cite{Furuta:2026eua}, non-unitary interface on lattice~\cite{Tang:2026ccf}, and exceptional points in holographic boundary CFTs~\cite{Maeda:2026awj}. It would be interesting to further explore the connections between these developments and our construction.

\begin{acknowledgments}
\emph{Acknowledgments.---}
I thank Ning Su and Yunqin Zheng for helpful discussions. Some of the computations were done and checked with the help of ChatGPT/Codex. I am supported by the Fujian Provincial Natural Science Foundation of China (No.~2025J01004) and the National Natural
Science Foundation of China (Grant No.~12247103).
\end{acknowledgments}

\bibliographystyle{apsrev4-2}
\bibliography{references}

\onecolumngrid
\clearpage
\begin{center}\textbf{End Matter}\end{center}
\twocolumngrid
\setcounter{equation}{0}
\renewcommand{\theequation}{E\arabic{equation}}
\renewcommand{\theHequation}{endmatter.\arabic{equation}}

\emph{An family of conformal interfaces between Ising and a non-unitary minimal model.---}
Our construction also applies when only one bulk theory is non-unitary. Let's consider the non-unitary minimal model $\mathcal T_{L}'=\mathcal M_{5,12}$ and $\mathcal T_R'={\rm Ising}$. Since both of them share an invertible $\mathbb Z_2$ symmetry, corresponding to the $\epsilon$ module with weight $\frac12$ in Ising, and $\mathcal V_{1,11}\in\mathcal M_{5,12}$, a natural auxiliary theory would be the diagonal $\mathbb Z_2$-orbifold of $\mathcal M_{5,12}\otimes\overline{\rm Ising}$. However the endpoint operator in $\mathcal V_{1,11}$ has conformal weight $\frac{15}{2}$. Therefore the condensation results in a spin-eight phantom current, which cannot be used to construct twisted Cardy branes in our setup. However, notice that there is an exceptional $E_6$ extension of $\mathcal M_{5,12}$ obtained by condensing $\mathcal V_{1,7}$ into its vacuum module, and $\mathcal V_{1,11}$ to $\mathcal V_{1,5}$. Therefore, in the folded picture for $\mathcal T_A'=\mathcal M_{5,12}\otimes\overline{\rm Ising}$, not only can we condense the phantom sector $\mathcal V_{1,11}\otimes\overline\epsilon$, but also two more modules $\mathcal V_{1,7}\otimes\overline{\one_{\rm Ising}}$ and $\mathcal V_{1,5}\otimes\overline{\epsilon}$. Therefore we can choose our auxiliary theory $\mathcal T_B'$ by condensing
\begin{align}
    \mathfrak a'=(V_{1,1}\oplus V_{1,7})\otimes\overline{\one_{\rm Ising}}\oplus (V_{1,5}\oplus V_{1,11})\otimes\overline{\epsilon}\,.
\end{align}
The endpoint of module $\mathcal V_{1,5}\otimes\overline{\epsilon}$ will then provide the spin-one phantom current. It is easy to show that the auxiliary theory $\mathcal T_B'$ is equivalent to 
\begin{align}
    \mathcal T_B'=\mathcal M_{2,5}\otimes \widehat{\mathfrak u}(1)_2
\end{align}
In this formulation, the spin-one phantom current can be identified with the affine $\widehat{\mathfrak u}(1)$ current, $Q=J_{\widehat{\mathfrak u}(1)}$. Using its zero mode $Q_0$, we can define a twisted Cardy brane $|b_\theta\rangle_{B'} \equiv e^{\ii \theta Q_0}\otimes\overline{\one}|b_0\rangle_{B'}$, where $|b_0\rangle_{B'}$ is the identity-labelled ordinary Cardy brane. Further using $\check{\mathcal I}_{B'\to A'}$, we end up with the pullback Cardy brane in $\mathcal T_A'$,
\begin{align}
    |\mathcal I_\theta\rangle_{A'}= \check{\mathcal I}_{B'\to A'}|b_\theta\rangle_{B'} 
\end{align}
The transmission rate of $\mathcal I_\theta$ can be computed as
\begin{align}
    \mathcal T(\theta)=-\frac{5}{17}\sin^2\theta\,,
\end{align}
generically nonzero for $\theta\neq 0$.

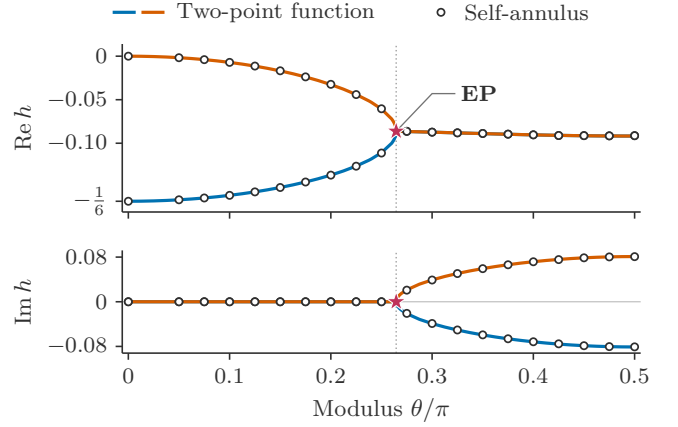
\begin{figure}[!t]
\centering
\begingroup%
\definecolor{EmEightBlue}{HTML}{0072B2}%
\definecolor{EmEightOrange}{HTML}{D55E00}%
\definecolor{EmEightInk}{HTML}{303030}%
\definecolor{EmEightStarColor}{HTML}{BF3159}%
\pgfdeclareplotmark{EmEightStar}{%
  \pgfpathmoveto{\pgfqpointpolar{90}{\pgfplotmarksize}}%
  \pgfpathlineto{\pgfqpointpolar{126}{0.381966\pgfplotmarksize}}%
  \pgfpathlineto{\pgfqpointpolar{162}{\pgfplotmarksize}}%
  \pgfpathlineto{\pgfqpointpolar{198}{0.381966\pgfplotmarksize}}%
  \pgfpathlineto{\pgfqpointpolar{234}{\pgfplotmarksize}}%
  \pgfpathlineto{\pgfqpointpolar{270}{0.381966\pgfplotmarksize}}%
  \pgfpathlineto{\pgfqpointpolar{306}{\pgfplotmarksize}}%
  \pgfpathlineto{\pgfqpointpolar{342}{0.381966\pgfplotmarksize}}%
  \pgfpathlineto{\pgfqpointpolar{18}{\pgfplotmarksize}}%
  \pgfpathlineto{\pgfqpointpolar{54}{0.381966\pgfplotmarksize}}%
  \pgfpathclose\pgfusepathqfillstroke%
}%
\def\EmEightEP{0.2645469924242915}%
\def\EmEightEPH{-0.08619852793408594}%
\pgfplotstableread[row sep=\\]{%
u rm rp im ip\\%
0 -0.1666666666666667 0 0 0\\%
0.0125 -0.1665628187640481 -0.0001067875853567256 0 0\\%
0.025 -0.1662508663852283 -0.0004279941167938074 0 0\\%
0.0375 -0.1657295667152884 -0.000966154987219837 0 0\\%
0.05 -0.1649967909546797 -0.001725510846471757 0 0\\%
0.0625 -0.1640494329219251 -0.002712035373532264 0 0\\%
0.075 -0.1628832697878102 -0.003933485768165232 0 0\\%
0.08749999999999999 -0.1614927614804213 -0.005399490074366465 0 0\\%
0.1 -0.1598707677516141 -0.007121693713320404 0 0\\%
0.1125 -0.1580081502476341 -0.009114000162462626 0 0\\%
0.125 -0.1558932081122309 -0.01139296066986033 0 0\\%
0.1375 -0.1535108639651063 -0.01397840081258506 0 0\\%
0.15 -0.1508414613221021 -0.01689442869081622 0 0\\%
0.1625 -0.1478589313435112 -0.02017107375441534 0 0\\%
0.175 -0.1445278841552892 -0.02384700983282593 0 0\\%
0.1875 -0.1407987570214532 -0.02797422966419915 0 0\\%
0.2 -0.1365991656259226 -0.03262655654388037 0 0\\%
0.2125 -0.1318171098121558 -0.03791630866607276 0 0\\%
0.225 -0.1262641322991487 -0.04403106253136492 0 0\\%
0.2375 -0.1195783027284856 -0.05133060330126277 0 0\\%
0.24 -0.1180460344639433 -0.05299158547228146 0 0\\%
0.25 -0.1108719017619795 -0.06069931983093393 0 0\\%
0.26 -0.1001076756487107 -0.07202527675139554 0 0\\%
0.26025 -0.09972369437786242 -0.07242363211438033 0 0\\%
0.2605 -0.09932814473951351 -0.07283356943359537 0 0\\%
0.26075 -0.09891994101965922 -0.0732561732211954 0 0\\%
0.261 -0.09849781813191556 -0.07369270638211425 0 0\\%
0.26125 -0.098060290132953 -0.07414465037862764 0 0\\%
0.2615 -0.09760560268298898 -0.07461374852853729 0 0\\%
0.26175 -0.09713171289620527 -0.07510200705408768 0 0\\%
0.262 -0.09663678148609346 -0.07561102391085134 0 0\\%
0.26225 -0.09610711306504327 -0.07615751972768128 0 0\\%
0.2625 -0.09555997155963419 -0.07671793752450544 0 0\\%
0.26275 -0.09497382218097995 -0.07731845805224663 0 0\\%
0.263 -0.09434433669694517 -0.07796243058588939 0 0\\%
0.26325 -0.0936608101390209 -0.07866048982838238 0 0\\%
0.2635 -0.09290673171833731 -0.0794291293016306 0 0\\%
0.26375 -0.09205479345987609 -0.08029565064893532 0 0\\%
0.264 -0.09105349974941714 -0.08131154659587329 0 0\\%
0.26425 -0.0897790875496844 -0.08260057865913746 0 0\\%
0.2644 -0.08871934587259366 -0.08366910033009635 0 0\\%
0.2645 -0.08762489810875194 -0.08476940470223331 0 0\\%
0.26455 -0.08619861604334393 -0.08619861604334393 -0.000361199051200321 0.000361199051200321\\%
0.2646 -0.08620008100269494 -0.08620008100269494 -0.001516381104459609 0.001516381104459609\\%
0.2647 -0.08620301187958015 -0.08620301187958015 -0.002576309007446979 0.002576309007446979\\%
0.26475 -0.0862044777942227 -0.0862044777942227 -0.00296755400509233 0.00296755400509233\\%
0.265 -0.0862118120794904 -0.0862118120794904 -0.004433024736398069 0.004433024736398069\\%
0.26525 -0.08621915412388308 -0.08621915412388308 -0.005522452900425297 0.005522452900425297\\%
0.2655 -0.08622650382346118 -0.08622650382346118 -0.006429904146674765 0.006429904146674765\\%
0.26575 -0.08623386109602206 -0.08623386109602206 -0.007224279414329931 0.007224279414329931\\%
0.266 -0.08624122587396779 -0.08624122587396779 -0.007939589811330115 0.007939589811330115\\%
0.26625 -0.08624859809979915 -0.08624859809979915 -0.00859559492190853 0.00859559492190853\\%
0.2665 -0.08625597772317301 -0.08625597772317301 -0.009204981045128405 0.009204981045128405\\%
0.26675 -0.08626336469892473 -0.08626336469892473 -0.009776467923106821 0.009776467923106821\\%
0.267 -0.08627075898570576 -0.08627075898570576 -0.01031635431923839 0.01031635431923839\\%
0.26725 -0.08627816054502388 -0.08627816054502388 -0.01082936600215328 0.01082936600215328\\%
0.2675 -0.08628556934055415 -0.08628556934055415 -0.01131915623638801 0.01131915623638801\\%
0.26775 -0.08629298533763516 -0.08629298533763516 -0.01178861843207325 0.01178861843207325\\%
0.268 -0.08630040850289522 -0.08630040850289522 -0.01224009054399573 0.01224009054399573\\%
0.26825 -0.08630783880397089 -0.08630783880397089 -0.01267549378312082 0.01267549378312082\\%
0.2685 -0.08631527620929227 -0.08631527620929227 -0.01309642973498677 0.01309642973498677\\%
0.26875 -0.08632272068791742 -0.08632272068791742 -0.01350425018290351 0.01350425018290351\\%
0.269 -0.08633017220940317 -0.08633017220940317 -0.01390010846459991 0.01390010846459991\\%
0.26925 -0.08633763074370343 -0.08633763074370343 -0.01428499800236724 0.01428499800236724\\%
0.2695 -0.08634509626108827 -0.08634509626108827 -0.01465978171741326 0.01465978171741326\\%
0.26975 -0.08635256873207921 -0.08635256873207921 -0.01502521483347515 0.01502521483347515\\%
0.27 -0.08636004812739681 -0.08636004812739681 -0.01538196279968033 0.01538196279968033\\%
0.275 -0.08651104567956819 -0.08651104567956819 -0.02129491319396171 0.02129491319396171\\%
0.2875 -0.08689906719701262 -0.08689906719701262 -0.03152240321442566 0.03152240321442566\\%
0.3 -0.08729900604776404 -0.08729900604776404 -0.03908885624505155 0.03908885624505155\\%
0.3125 -0.08770665886071834 -0.08770665886071834 -0.04530404849869836 0.04530404849869836\\%
0.325 -0.08811753825982677 -0.08811753825982677 -0.05063013399702129 0.05063013399702129\\%
0.3375 -0.08852696235767328 -0.08852696235767328 -0.05529084615936462 0.05529084615936462\\%
0.35 -0.08893014606835244 -0.08893014606835244 -0.05941262539516995 0.05941262539516995\\%
0.3625 -0.08932229093238325 -0.08932229093238325 -0.06307391589991601 0.06307391589991601\\%
0.375 -0.08969867111544461 -0.08969867111544461 -0.06632651480375787 0.06632651480375787\\%
0.3875 -0.09005471379993894 -0.09005471379993894 -0.06920616995542599 0.06920616995542599\\%
0.4 -0.09038607270385006 -0.09038607270385006 -0.07173836366075098 0.07173836366075098\\%
0.4125 -0.0906886939343943 -0.0906886939343943 -0.07394169496894128 0.07394169496894128\\%
0.425 -0.09095887379409529 -0.09095887379409529 -0.07582996237324088 0.07582996237324088\\%
0.4375 -0.09119330848931874 -0.09119330848931874 -0.07741349675316431 0.07741349675316431\\%
0.45 -0.091389135939188 -0.091389135939188 -0.07870003859775483 0.07870003859775483\\%
0.4625 -0.09154397004641245 -0.09154397004641245 -0.07969532560303662 0.07969532560303662\\%
0.475 -0.09165592787651015 -0.09165592787651015 -0.0804034884956681 0.0804034884956681\\%
0.4875 -0.09172365020747252 -0.09172365020747252 -0.08082731439747254 0.08082731439747254\\%
0.5 -0.09174631586958404 -0.09174631586958404 -0.08096841402838978 0.08096841402838978\\%
}\EmEightSpectrum%
\pgfplotstableread[row sep=\\]{%
u rm rp im ip\\%
0 -0.1666666666666667 0 0 0\\%
0.05 -0.164974114184244 -0.001756581440813975 0 0\\%
0.075 -0.1628848126931308 -0.003989045800708879 0 0\\%
0.1 -0.1598724317133965 -0.007118510251660723 0 0\\%
0.125 -0.1558982831399238 -0.01134544617094891 0 0\\%
0.15 -0.1508537762762531 -0.01678368501471349 0 0\\%
0.175 -0.1445489781082497 -0.02375923185988844 0 0\\%
0.2 -0.1366469223454286 -0.03233243661429354 0 0\\%
0.225 -0.1263761973597767 -0.04409674016053605 0 0\\%
0.25 -0.1112715695187839 -0.060400520790939 0 0\\%
0.275 -0.08634367144255604 -0.08634161003959087 0.02081053455644251 -0.02081173369767799\\%
0.3 -0.08728167884918495 -0.08728650614670799 0.0389545396408811 -0.03893835890600782\\%
0.325 -0.08805013619131591 -0.08804877494924609 0.05053642302331493 -0.0505399961947748\\%
0.35 -0.08879494693368857 -0.0887948099606659 0.05930872504175703 -0.05930854903210109\\%
0.375 -0.08959061939220012 -0.0895900208966921 0.06626959976855815 -0.06627376240830936\\%
0.4 -0.09034836960512266 -0.09034709492477372 0.07165632689915598 -0.07165379529185707\\%
0.425 -0.0911738686408701 -0.09110631461182457 0.07553107310537951 -0.0753436728994491\\%
0.45 -0.09127767697799261 -0.09127765786429268 0.07872122466764414 -0.07872155271173736\\%
0.475 -0.09150788513449792 -0.09150788656353465 0.08035759157243719 -0.08035679906085524\\%
0.5 -0.09133162354836726 -0.09133251063790482 0.08080330748037001 -0.08080320864058746\\%
}\EmEightAnnulus%
\begin{tikzpicture}[x=1in,y=1in]
\path[use as bounding box] (0,0) rectangle (3.375,2.20);
\pgfplotsset{
  em eight axes/.style={
    scale only axis,width=2.70in,
    axis lines*=left,axis line style={EmEightInk,line width=.60pt},
    xmin=-.006,xmax=.506,xtick={0,.1,.2,.3,.4,.5},scaled ticks=false,
    tick align=outside,major tick length=2.5pt,
    tick style={EmEightInk,line width=.60pt},
    tick label style={font=\fontsize{8}{9.6}\selectfont,text=EmEightInk},
    label style={font=\fontsize{9}{10.8}\selectfont,text=EmEightInk},
    ylabel style={at={(axis description cs:-.16,.5)},anchor=south},
    clip mode=individual,axis on top=true,
  },
  em eight curve/.style={line width=1.20pt,no marks},
  em eight annulus/.style={only marks,mark=*,mark size=1.35pt,
    mark options={draw=EmEightInk,fill=white,line width=.65pt}},
  em eight star/.style={only marks,mark=EmEightStar,mark size=3.8pt,
    mark options={draw=white,fill=EmEightStarColor,line width=.40pt}},
}
\begin{axis}[
  em eight axes,at={(.61in,1.075in)},anchor=south west,height=.875in,
  ymin=-.179,ymax=.014,ytick={-.166666666666667,-.1,-.05,0},
  yticklabels={$-\frac16$,$-0.10$,$-0.05$,$0$},
  xticklabels=\empty,ylabel={$\operatorname{Re}h$},
]
\draw[black!35,line width=.55pt,densely dotted]
  (axis cs:\EmEightEP,-.179)--(axis cs:\EmEightEP,.014);
\addplot[em eight curve,EmEightBlue] table[x=u,y=rm]{\EmEightSpectrum};
\addplot[em eight curve,EmEightOrange] table[x=u,y=rp]{\EmEightSpectrum};
\addplot[em eight annulus] table[x=u,y=rm]{\EmEightAnnulus};
\addplot[em eight annulus] table[x=u,y=rp]{\EmEightAnnulus};
\draw[black!55,line width=.5pt]
  (axis cs:\EmEightEP,\EmEightEPH)--(axis cs:.300,-.043)--(axis cs:.323,-.043);
\node[anchor=west,inner sep=0pt,font=\fontsize{8}{9.6}\selectfont\bfseries,
  text=EmEightInk] at (axis cs:.328,-.043) {EP};
\addplot[em eight star] coordinates {(\EmEightEP,\EmEightEPH)};
\end{axis}
\begin{axis}[
  em eight axes,at={(.61in,.34in)},anchor=south west,height=.535in,
  ymin=-.092,ymax=.092,ytick={-.08,0,.08},
  yticklabels={$-0.08$,$0$,$0.08$},
  ylabel={$\operatorname{Im}h$},xlabel={Modulus $\theta/\pi$},
  xlabel style={at={(axis description cs:.5,-.35)},anchor=north},
]
\draw[black!25,line width=.45pt] (axis cs:-.006,0)--(axis cs:.506,0);
\draw[black!35,line width=.55pt,densely dotted]
  (axis cs:\EmEightEP,-.092)--(axis cs:\EmEightEP,.092);
\addplot[em eight curve,EmEightBlue] table[x=u,y=im]{\EmEightSpectrum};
\addplot[em eight curve,EmEightOrange] table[x=u,y=ip]{\EmEightSpectrum};
\addplot[em eight annulus] table[x=u,y=im]{\EmEightAnnulus};
\addplot[em eight annulus] table[x=u,y=ip]{\EmEightAnnulus};
\addplot[em eight star] coordinates {(\EmEightEP,0)};
\end{axis}
\draw[EmEightBlue,line width=1.20pt] (.56,2.115)--(.68,2.115);
\draw[EmEightOrange,line width=1.20pt] (.71,2.115)--(.83,2.115);
\node[anchor=west,inner sep=0pt,font=\fontsize{8}{9.6}\selectfont,text=EmEightInk]
  at (.89,2.115) {Two-point function};
\draw[EmEightInk,fill=white,line width=.65pt] (2.27,2.115) circle[radius=1.35pt];
\node[anchor=west,inner sep=0pt,font=\fontsize{8}{9.6}\selectfont,text=EmEightInk]
  at (2.39,2.115) {Self-annulus};
\end{tikzpicture}%
\endgroup%

\caption{Spectral collision between the $h=0$ and $-\frac{1}{6}$ branches}
\label{fig:em-spectrum}
\end{figure}

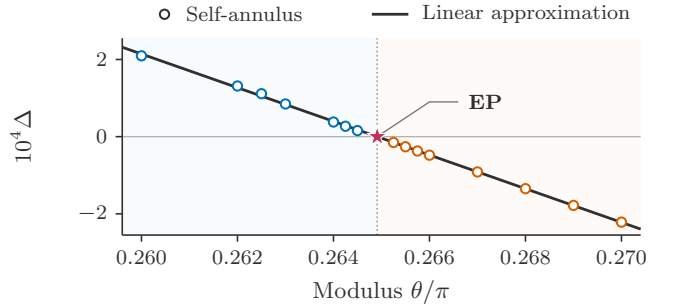
\begin{figure}[!t]
\centering
\begingroup%
\definecolor{EmEightBlue}{HTML}{0072B2}%
\definecolor{EmEightOrange}{HTML}{D55E00}%
\definecolor{EmEightInk}{HTML}{303030}%
\definecolor{EmEightStarColor}{HTML}{BF3159}%
\pgfdeclareplotmark{EmEightStar}{%
  \pgfpathmoveto{\pgfqpointpolar{90}{\pgfplotmarksize}}%
  \pgfpathlineto{\pgfqpointpolar{126}{0.381966\pgfplotmarksize}}%
  \pgfpathlineto{\pgfqpointpolar{162}{\pgfplotmarksize}}%
  \pgfpathlineto{\pgfqpointpolar{198}{0.381966\pgfplotmarksize}}%
  \pgfpathlineto{\pgfqpointpolar{234}{\pgfplotmarksize}}%
  \pgfpathlineto{\pgfqpointpolar{270}{0.381966\pgfplotmarksize}}%
  \pgfpathlineto{\pgfqpointpolar{306}{\pgfplotmarksize}}%
  \pgfpathlineto{\pgfqpointpolar{342}{0.381966\pgfplotmarksize}}%
  \pgfpathlineto{\pgfqpointpolar{18}{\pgfplotmarksize}}%
  \pgfpathlineto{\pgfqpointpolar{54}{0.381966\pgfplotmarksize}}%
  \pgfpathclose\pgfusepathqfillstroke%
}%
\def\EmEightDeltaZero{0.2649103104553387}%
\def\EmEightDeltaSlope{-0.0436396557139258}%
\begin{tikzpicture}[x=1in,y=1in]
\path[use as bounding box] (0,0) rectangle (3.375,1.58);
\begin{axis}[
  at={(.61in,.34in)},anchor=south west,
  width=2.70in,height=1.025in,scale only axis,
  axis lines*=left,axis line style={EmEightInk,line width=.60pt},
  xmin=.2596,xmax=.2704,ymin=-2.55,ymax=2.55,
  xtick={.26,.262,.264,.266,.268,.27},ytick={-2,0,2},
  xticklabel style={/pgf/number format/fixed,/pgf/number format/precision=3,
    /pgf/number format/fixed zerofill},
  scaled ticks=false,tick align=outside,major tick length=2.5pt,
  tick style={EmEightInk,line width=.60pt},
  tick label style={font=\fontsize{8}{9.6}\selectfont,text=EmEightInk},
  label style={font=\fontsize{9}{10.8}\selectfont,text=EmEightInk},
  xlabel={Modulus $\theta/\pi$},ylabel={$10^4\Delta$},
  xlabel style={at={(axis description cs:.5,-.185)},anchor=north},
  ylabel style={at={(axis description cs:-.16,.5)},anchor=south},
  clip mode=individual,axis on top=true,
]
\path[fill=EmEightBlue!3!white]
  (rel axis cs:0,0) rectangle (axis cs:\EmEightDeltaZero,2.55);
\path[fill=EmEightOrange!3!white]
  (axis cs:\EmEightDeltaZero,-2.55) rectangle (rel axis cs:1,1);
\draw[black!30,line width=.5pt] (axis cs:.2596,0)--(axis cs:.2704,0);
\draw[black!40,densely dotted,line width=.6pt]
  (axis cs:\EmEightDeltaZero,-2.55)--(axis cs:\EmEightDeltaZero,2.55);
\addplot[EmEightInk,line width=1.05pt,no marks,domain=.2596:.2704,samples=2]
  {10000*\EmEightDeltaSlope*(x-\EmEightDeltaZero)};
\addplot[only marks,mark=*,mark size=1.75pt,
  mark options={draw=EmEightBlue,fill=white,line width=.75pt}]
  coordinates {
(0.26,2.094610852952811)
(0.262,1.313416343191268)
(0.2625,1.113008308058584)
(0.263,0.8466828982339242)
(0.264,0.3791398731759494)
(0.26425,0.2695062388710256)
(0.2645,0.1571667220074274)
};
\addplot[only marks,mark=*,mark size=1.75pt,
  mark options={draw=EmEightOrange,fill=white,line width=.75pt}]
  coordinates {
(0.26525,-0.1509143090293813)
(0.2655,-0.2619760464394301)
(0.26575,-0.3714085472221282)
(0.266,-0.4803190939065985)
(0.267,-0.9145026821484777)
(0.268,-1.348142074137775)
(0.269,-1.781724988154132)
(0.27,-2.215340597969734)
};
\draw[black!55,line width=.5pt]
  (axis cs:\EmEightDeltaZero,0)--(axis cs:.266,.9)--(axis cs:.2666,.9);
\node[anchor=west,inner sep=0pt,font=\fontsize{7.8}{9.4}\selectfont\bfseries,text=EmEightInk]
  at (axis cs:.2668,.9) {EP};
\addplot[only marks,mark=EmEightStar,mark size=3.8pt,
  mark options={draw=white,fill=EmEightStarColor,line width=.4pt}]
  coordinates {(\EmEightDeltaZero,0)};
\end{axis}
\draw[EmEightInk,fill=white,line width=.75pt] (.83,1.49) circle[radius=1.75pt];
\node[anchor=west,inner sep=0pt,font=\fontsize{8}{9.6}\selectfont,text=EmEightInk]
  at (.94,1.49) {Self-annulus};
\draw[EmEightInk,line width=1.05pt] (1.91,1.49)--(2.10,1.49);
\node[anchor=west,inner sep=0pt,font=\fontsize{8}{9.6}\selectfont,text=EmEightInk]
  at (2.17,1.49) {Linear approximation};
\end{tikzpicture}%
\endgroup%

\caption{The squared half-gap
$\Delta=(h_+-h_-)^2/4$ from the self-annulus near the collision.
The linear fit has a linear simple zero at $\theta_*/\pi\simeq0.2649$, consistent with the correlator result.}
\label{fig:em-delta}
\end{figure}

With $\mathcal I_\theta$ at hand, we can compute again its self-annulus partition function and extract its spectrum. We find that the lowest spectral collision happens between the branch of $h=-\frac16$ and $h=0$ at $\theta=0$. Again we investigate the logarithmic scaling behavior by choosing a probe $\Phi'=\phi_{1,2}^{\mathcal M_{5,12}}\otimes\overline{\one_{\rm Ising}}$ with weight $h_{\Phi'}=-\frac{3}{16}$, whose bulk-boundary OPE produces the two target boundary operators via the $t$-channel. The numerical results are summarized below: FIG~\ref{fig:em-spectrum} compares the self-annulus spectrum with the disk correlator of $\Phi'$. The two branches starting at $h=-1/6$ and $h=0$ collide at $\theta_*/\pi\simeq0.2646$, $h_*\simeq-0.0862$, and become
$h\simeq-0.09175\pm0.08097\,\mathrm i$ at $\theta=\pi/2$. At the collision the correlator contains 
\begin{align}
 A_{\Phi'}\mathcal F_{h_*}
 +K_{\Phi'}\left.\partial_h\mathcal F_h\right|_{h=h_*}\,,
 \qquad K_{\Phi'}\simeq-0.3930\,,
 \label{eq:em-logblock}
\end{align}
In addition, the annulus data support a simple zero of the squared half-gap, see FIG.~\ref{fig:em-delta}.

\emph{Truncated crossing equation.---}
In the $s$ channel, $G_\lambda(z)$ in Eq.~\eqref{eq:s-channel_G} is determined by the bulk OPE of $\phi_{1,3}$ and its contraction with the boundary state $\mathcal I_\lambda$. Using the null relations of $\phi_{1,3}$, one can compute it order by order in $z$. For $\lambda=0$, the crossing equation~\eqref{eq:crossing} can be worked out exactly. We express the result in terms of the $\mathcal M_{4,7}$ Virasoro blocks $\cF_h^{\lambda=0}(z)$, with central charge $c=-13/14$, four external weights $1/7$ and intermediate weight $h\in\left\{0,\frac17,\frac{10}{7}\right\}$. These blocks are normalized as $\cF_h^{\lambda=0}(z)=z^{h-2/7}\left(1+\mathcal O(z)\right)$ and take the explicit form
\begin{align}
\cF_h^{\lambda=0}(z)
&=z^{h-2/7}(1-z)^{-h/2}\\
&\quad\times{}_3F_2\!\left(
\begin{matrix}
\frac h2-\frac27,\;\frac h2-\frac17,\;\frac h2+\frac87\\
1+\frac{h-h'}2,\;1+\frac{h-h''}2
\end{matrix}
;\frac{-z^2}{4(1-z)}\right)\,,\notag
\end{align}
where $(h,h',h'')$ is any permutation of $(0,1/7,10/7)$. The crossing equations~\eqref{eq:crossing} then reads
\begin{align}
  &G_0(z)=\cF_0^{\lambda=0}(z)\\[3pt]
  &=a_0F_0^{\lambda=0}(1-z)+a_{\frac17}\cF_{\frac17}^{\lambda=0}(1-z)+a_{\frac{10}{7}}\cF_{\frac{10}{7}}^{\lambda=0}(1-z)\,,\notag
\end{align}
with $a_0=-\frac{\Gamma(\frac27)\Gamma(\frac57)}{\Gamma(\frac37)\Gamma(\frac47)}$, $a_{\frac17}=\frac{28\Gamma(\frac27)^2\Gamma(\frac47)\Gamma(\frac67)^2}{\Gamma(\frac17)^2\Gamma(\frac37)^2\Gamma(\frac57)}$ and $a_{\frac{10}{7}}=\frac{2\Gamma(\frac27)^2}{15\Gamma(\frac17)\Gamma(\frac37)}$. For $\lambda\neq0$, we write the ansatz
\begin{align}
 G_\lambda(z)
 =\mu_-(\lambda)\cF_{h_-(\lambda)}(\eta)
 +\mu_+(\lambda)\cF_{h_+(\lambda)}(\eta)
+\mathcal R_\lambda(z)\,,
 \label{eq:em-crossing-fit}
\end{align}
where $h_\pm$ are the two target dimensions, $\mu_\pm$ encode the corresponding OPE coefficients, and $\mathcal R_\lambda$ contains all other boundary contributions. Both dimensions and coefficients will be determined by the fit.

To approximate $\mathcal R_\lambda$, we choose a basis motivated by perturbation around $\lambda=0$, where the boundary channel consists of the three $\mathcal M_{4,7}$ blocks $\mathcal F_h^{\lambda=0}$ with $h=0,\frac17,\frac{10}{7}$. We first decompose them into blocks $\cF_h$ of the total Virasoro algebra of $\mathfrak A$, 
\begin{align}
 \cF_h^{\lambda=0}(\eta)
 &=\sum_{n=0}^{\infty}a_{h,n}\cF_{h+2n}(\eta)\,,
 \quad h\in\left\{0,\frac17,\frac{10}{7}\right\}\,,
\end{align}
We exclude the leading blocks at weights $0$ and $\frac17$, since their evolving contributions will be fitted separately as our target pair. The remaining blocks provide reference functions for $\mathcal R_\lambda$. For such a block $\mathcal F_h$, as $\lambda$ varies, both its coefficients and weights can change. We therefore use both $\mathcal F_h$ and $\partial_h\mathcal F_h$ to approximate $\mathcal R_\lambda$. On the other hand, the marginal spin-one phantom current contains $\phi_{1,2}$ of $\mathcal M_{4,7}$. Its fusion with the original boundary operator $\phi_{1,1}$, $\phi_{1,3}$ and $\phi_{1,5}$ gives $\phi_{1,2}$, $\phi_{1,4}$ and $\phi_{1,6}$ in $\mathcal M_{4,7}$. They will give contributions of $\mathcal F_{h'+2n}$ for $h'=1,\,\frac{12}{7}$ and $\frac{25}{7}$. Overall the $\mathcal R_\lambda$ approximation is
\begin{align}
 \mathcal R_\lambda(z)
 &\simeq\mathcal R_0(z)
 +\sum_{a\in\mathcal S}
 \left(
 u_a(\lambda)\cF_a(\eta)
 +v_a(\lambda)\left.\partial_h\cF_h(\eta)\right|_{h=a}
 \right)
 \notag\\
 &\quad+\sum_{b\in\mathcal N}w_b(\lambda)\cF_b(\eta)\,,
 \label{eq:em-crossing-remainder}
\end{align}
with
\begin{align}
 \mathcal S
 &=\left(\bigcup_{n=0}^{4}
 \left\{2n,\frac17+2n,\frac{10}{7}+2n\right\}\right)
 \setminus\left\{0,\frac17\right\}\,,
 \notag\\
 \mathcal N
 &=\bigcup_{n=0}^{1}
 \left\{1+2n,\frac{12}{7}+2n,\frac{25}{7}+2n\right\}\,,
 \label{eq:em-crossing-basis}
\end{align}
where $\mathcal R_0=a_{\frac{10}{7}}\cF_{\frac{10}{7}}^{\lambda=0}(1-z)$ as a $\lambda=0$ benchmark.

we compute $G_\lambda(z)$ up to $\mathcal O(z^{20})$ while retaining their coefficients with full dependence on $\lambda$. The coefficient of $\lambda^2$ is also computed up to  $\mathcal O(z^{32})$. For numerical evaluation, we factor out the standard elliptic prefactor and re-expand the remaining series in the bulk elliptic nome $q(z)$ order 40~\cite{Perlmutter:2015iya}, using 135 digit arithmetic. At each modulus $\lambda$, we evaluate $G_\lambda$ at 45 points in $0.3\leq z\leq0.6$. For trial $h_\pm$, we determine all coefficients by linear least squares, and then optimize $h_\pm$ by minimizing the fitting residual. Near the collision we use their mean and squared half difference, see Eq.~\eqref{eq:m_Delta}, together with the redefiniation of the blocks $\mathcal E_{m,\Delta}=\frac{\mathcal F_{m+\delta}+\mathcal F_{m-\delta}}{2}$ and $\mathcal D_{m,\Delta}=\frac{\mathcal F_{m+\delta}-\mathcal F_{m-\delta}}{2\delta}$. The latter tends to a weight derivative when the dimensions coincide. We test the approximation by varying the retained functions, the order of $G_\lambda(z)$ and the fitting interval separately.

\emph{Modular data.---}
We list the modular data of the two bulk theories defined in Eq.~\eqref{eq:osp}. One can also obtain them using the modular data of (non-)unitary minimal models and the orbifold dualities in Eq.~\eqref{eq:duality}. For the $\widehat{\mathfrak {osp}}(1|2)_k$ coset construction, $\mathcal T_L$, $\mathcal T_R$ and the auxiliary $\mathcal T_B$ have respectively $12$, $12$ and $16$ primaries, with central charge $c_L=-\frac37$, $c_R=\frac{8}{35}$ and $c_B=-\frac15$. Their primaries can be labelled by 
\begin{align}
    &L_{r;\,m}\in\frac{\widehat{\mathfrak {osp}}(1|2)_2}{\widehat{\mathfrak u}(1)_2}\,,\qquad 
    R_{e_1,e_2;\,r}\in \frac{\widehat{\mathfrak {osp}}(1|2)_1^{\otimes2}}{\widehat{\mathfrak {osp}}(1|2)_2}\,,\notag\\[4pt]
    &B_{e_1,e_2;\,r}\in\frac{\widehat{\mathfrak {osp}}(1|2)_1^{\otimes2}}{\widehat{\mathfrak u}(1)_2}\,,
\end{align}
with $r\in\{1,3,5\}$, $e_1,\,e_2\in\{1,3\}$ and $m\in\mathbb Z_4$. Here $r$ is the $\widehat{\mathfrak {osp}}(1|2)_2$ label, $e_i$ are the $\widehat{\mathfrak {osp}}(1|2)_1$ labels, and $m$ is the $\widehat{\mathfrak u}(1)_2$ charges. The grade parafermion branching and the diagonal coset extension are described in~\cite{https://doi.org/10.48550/arxiv.1706.00242,https://doi.org/10.48550/arxiv.2509.14654}. The phantom sectors in Eq.~\eqref{eq:phantom_sectors} are given by
\begin{align}
    &\mathcal X_\one=L_{1;0}\otimes\overline{R}_{0,0;1}\,,\quad
    \mathcal X_x=L_{3;0}\otimes\overline{R}_{0,0;3}\,,\quad\notag\\[3pt]
    &\mathcal X_y=L_{5;0}\otimes\overline{R}_{0,0;5}\,.
\end{align}
We list all their conformal weights below
\begin{align}
 &\begin{array}{c|ccc}
 h^{L}_{r;\,m} & r=1\quad&r=3\quad&r=5\\ \hline
 m=0&0&\frac17&\frac37\\
 m=1,3&\frac78&\frac{1}{56}&\frac{17}{56}\\
 m=2&\frac12&\frac{9}{14}&\!\!\!\!-\frac{1}{14}
 \end{array}\,,\notag
 \\[3pt]
 &\begin{array}{c|ccc}
 h^{R}_{e_1,e_2;\,r}& e_1+e_2=2\quad\quad&4\quad\quad&6\\ \hline
 r=1&0&\frac{11}{5}\quad\quad&\frac 25\\
 r=3&\frac{13}{7}&\frac{2}{35}\quad\quad&\frac{9}{35}\\
 r=5&\frac47&\frac{27}{35}\quad\quad&\!\!\!\!-\frac{1}{35}
 \end{array}\,,
 \label{eq:em-weights}
 \\[3pt]
 &\begin{array}{c|ccc}
 h^B_{m,e_1,e_2}& e_1+e_2=2\quad\quad&4\quad\quad&6\\ \hline
 m=0&0&\frac15\quad\quad&\frac25\\
 m=1,3&\frac78&\frac{3}{40}\quad\quad&\frac{11}{40}\\
 m=2&\frac12&\frac{7}{10}\quad\quad&\!\!\!\!\!-\frac{1}{10}
 \end{array}\,,\notag
\end{align}
where $m=1,3$ and $(e_1,e_2)=(1,3),(3,1)$ denote distinct primaries. Now define the matrices,
\begin{align}
 D_{rs}&=\frac2{\sqrt7}\sin\frac{4\pi rs}{7}\,,
 \qquad
 G_{mn}=\frac{(-1)^{m+n}}2e^{\mathrm i\pi mn/2}\,,
 \notag\\
 F_{ef}&=\frac2{\sqrt5}
 \sin\frac{3\pi ef}5\,.
 \label{eq:em-factors}
\end{align}
Using them, the full modular matrices of $\mathcal T_L$, $\mathcal T_R$, and $\mathcal T_B$ read as follows:
\begin{align}
 &S^{L}_{(r;\,m),(s;\,n)}=D_{rs}G_{mn}\,,\ 
 S^{R}_{(e_1,e_2;\,r),(f_1,f_2;\,s)}=D_{rs}F_{e_1f_1}F_{e_2f_2}\,,
 \notag\\[5pt]
 &S^B_{(e_1,e_2;\,m),(f_1,f_2;\,n)}
 =G_{mn}F_{e_1f_1}F_{e_2f_2}\,.
\label{eq:em-ospS}
\end{align}

\end{document}